    \def\parag#1{\vspace*{5mm}\noindent \parbox[t]{14cm}{\Large\bf #1}}

    \def\rp{r{\lower .3mm\hbox{\tiny '}}}
    \def\rpp{r{\lower .3mm\hbox{\tiny ''}}}

 \def\dup{\6\hspace{-1.6mm}\lower 1.2mm\hbox{\bf .}\,}
 \def\zup{ z\hspace{-1.6mm}\lower 1.2mm\hbox{\bf .}\,}

     \font\sm=cmbx5    % fonts dick, duenn s. Wurzel etc.
     \def\ft{\footnotesize}

     \def\glqq{``}  \def\grqq{''}
     \def\ep{\epsilon}
     \def\anke{$\; - \;$}

   \def\({\left(} \def\){\right)}  
   \def\lk{\,\left[ \,} \def\rk{\,\right] \,} 
   \def\rki{\,\right]}  \def\lb{\left\{} \def\rb{\right\}}

   \font\dick=cmbx12  \font\duenn=cmbx10 
   \def\dop{\duenn \raise0.6pt\hbox to 0.2pt{: \hss}}
   \def\dep{\duenn \raise0.6pt\hbox to 0.3pt{\hss :}}

   \def\cl#1{{\cal #1}}    \def\ct#1{\nz\centerline{#1}}
   \def\schl#1{\widetilde{#1}}  \def\ov#1{\overline{#1}}   
   \def\dis{\displaystyle}  \def\0{\over } \def\6{\partial }

   \def\gll{\; {\dop} =}   \def\glr{= {\dep} \;}
   \def\ln{{\rm ln}}   
   \def\Tr{\hbox{Tr}}  \def\Sp{\hbox{Sp}}

   \let\a=\alpha  \let\b=\beta    \let\d=\delta  
       \let\o=\omega  
   \let\s=\sigma      \let\e=\varepsilon  
   \let\ph=\varphi   
       \let\L=\Lambda  
   \let\O=\Omega
      
   \def\eq#1{(\ref{#1})}        \def\nonu{\nonumber}
   \def\be#1{\begin{equation} \label{#1}}
   \def\ben{\begin{equation}}   \def\ee{\end{equation}}
   \def\bea#1{\begin{eqnarray} \label{#1}}
   \def\bean{\begin{eqnarray}}  \def\eea{\end{eqnarray}}
   \let\thq=\theequation
   \def\pfeil{_\rightharpoonup}  \def\leer{\phantom{a}}
   \def\opf{\buildrel \pfeil \over \leer}
   \def\jvv{j \lower0.4pt\hbox to 2pt{\hss $\opf$}} 
   \def\jv{j \lower0.2pt\hbox to 1.4pt{\hss $\opf$}} 
   \def\ivv{i \lower0.4pt\hbox to 2pt{\hss $\opf$}} 
   \def\iv{i \lower0.2pt\hbox to 1.4pt{\hss $\opf$}} 
   \def\hq{h \raise0.2pt\hbox to 0.4pt{\hss $^-$}}   
   \def\vk#1{\hbox{$\buildrel           \pfeil \over #1$}}
   \def\vkk#1{\hbox{$\buildrel   \;     \pfeil \over #1$}}
   \def\vkkk#1{\hbox{$\buildrel  \, \;  \pfeil \over #1$}}
   \def\grpf{\displaystyle  _\rightharpoonup}
   \def\vg#1{\hbox{$\buildrel       \grpf \over #1$}}
   \def\vgg#1{\hbox{$\buildrel  \;  \grpf \over #1$}}
\def\fzz{f} \def\bzz{b} \def\dzz{d} \def\gzz{g} \def\hzz{h} 
\def\jzz{j} \def\kzz{k} \def\lzz{l} \def\mzz{m} \def\wzz{w} 
\def\tzz{t} \def\izz{i} 
\def\bezz{\beta} \def\dezz{\delta} \def\xizz{\xi} 
\def\pszz{\psi}  \def\vthzz{\vartheta}
\def\uph{ \! \mathop{\vphantom{a}} } \def\dph{ \vphantom{a} }
\def\vc#1{\def\tast{\noexpand#1} \def\test{#1}
\ifcat\tast\bzz 
\ifx\test\fzz \vkkk f \uph \else  \ifx\test\bzz \vkk b \uph \else
\ifx\test\dzz \vkkk d \uph \else  \ifx\test\gzz \vkk g \dph \else
\ifx\test\hzz \vkk h \uph \else   \ifx\test\izz \ivv \dph \else 
\ifx\test\jzz \jvv \dph \else     \ifx\test\kzz \vkk k \uph \else
\ifx\test\lzz \vkk l \uph \else   \ifx\test\tzz \vkk t \uph \else
\ifx\test\mzz \vg m \dph \else    \ifx\test\wzz \vg w \dph \else 
\ifnum \lq#1<91 \vgg #1 \uph \else \vk #1 \dph 
  \fi \fi \fi \fi \fi \fi \fi \fi \fi \fi \fi \fi \fi  \else
\ifx\test\bezz \vkk \beta \uph \else  
    \ifx\test\pszz \vkk \psi \dph \else 
\ifx\test\dezz \vkk \delta \uph \else 
    \ifx\test\xizz \vkk \xi \uph \else
\ifx\test\vthzz \vkk \vartheta \uph \else \vk #1 \dph
  \fi \fi \fi \fi \fi \fi }
\def\abst#1{\def\tast{\noexpand#1} \ifcat\tast\bzz 
    \ifnum \lq#1<91 \; \else \, \fi      \else \, \fi}
\def\absv#1{\def\tast{\noexpand#1} \def\test{#1}
\ifcat\tast\bzz \ifx\test\fzz \, \; \else  
    \ifx\test\dzz \, \; \else
\ifx\test\bzz \, \else \ifx\test\gzz \, \else 
       \ifx\test\hzz \, \else
\ifx\test\kzz \, \else \ifx\test\lzz \, \else 
       \ifx\test\tzz \, \else 
\ifnum \lq#1<91 \; \else   
  \fi \fi \fi \fi \fi \fi \fi \fi \fi \else
\ifx\test\bezz \, \else \ifx\test\pszz \, \else 
       \ifx\test\dezz \, \else
\ifx\test\xizz \, \else \ifx\test\vthzz \, \else 
   \fi \fi \fi \fi \fi \fi }
\def\vphan#1{\def\tast{\noexpand#1} \def\test{#1}
\ifcat\tast\bzz \ifx\test\fzz  \uph \else 
      \ifx\test\bzz  \uph \else
\ifx\test\dzz  \uph \else  \ifx\test\hzz  \uph \else
\ifx\test\kzz  \uph \else  \ifx\test\lzz  \uph \else
\ifx\test\tzz  \uph \else  \ifnum \lq#1<91 \uph \else \dph 
  \fi \fi \fi \fi \fi \fi \fi \fi \else
\ifx\test\bezz  \uph \else  \ifx\test\dezz  \uph \else
\ifx\test\xizz  \uph \else  \ifx\test\vthzz \uph \else \dph
  \fi \fi \fi \fi \fi }

  \def\pubox{\dick _{\raise 1pt\hbox{.}} }
  \def\ppubox{\dick _{\raise 1pt\hbox{..}} }

  \def\vcsm#1{ \def\sm{\raise 1.6pt\hbox to 5pt{\hss $_#1$}} 
               {\buildrel \pfeil \over \sm} \>\!\! }

   \def\abpfeil{ \raise 1pt\hbox{ $_\vee$ \hskip -6.8pt
          \vrule depth 0.6pt height 4.6pt width 0.3pt} \;\; }

\def\nz{$ $\\}  
\def\ft{\footnotesize}   

\def\unt#1#2#3#4{\hbox{
  \vrule height .2cm width 0.4pt  depth -.5pt
  \vrule height .5pt width #1cm   depth 0pt
  \vrule height .2cm width 0.3pt  depth -.5pt
  \hskip -#1cm \hskip #3cm  \vrule width 0.3pt 
  height 0pt depth #2cm \lower #2cm\hbox{\lower .14cm\hbox{$
  \!\!\!\,\dis = #4$}}} \nonu }
    \def\2{\hbox{${1\02}$}}

 \def\gef{\lower.5pt\hbox{$^\angle\hskip -.245cm
         $\raise .7pt\hbox{${}_{^\swarrow}$}}}
 \def\Sp{{\rm \,Tr}}
 \def\folg{\,\curvearrowright\,}

\def\glo#1{ & = \hspace{-.17cm}
            \vrule width 0.12pt height .06cm depth #1mm 
            \hspace{.12cm} & }
\def\glu#1{ \nonu \\[-.5cm] & = \hspace{-.17cm}
          \vrule width 0.12pt height #1mm depth -.14cm
           \hspace{.12cm} & }
\def\gluo#1#2{ \nonu \\[-.5cm] & \hspace{.005cm}
            = \hspace{-.177cm}
           \vrule width 0.12pt height #1mm depth -.14cm
           \hspace{-.003cm}
           \vrule width 0.12pt height .06cm depth #2mm
           \hspace{.12cm} & }

\def\krumm#1#2#3#4{\hspace*{.2mm}\unitlength 1cm
  \raise .26mm\hbox{\begin{picture}(.01,.2)
  \put(-.15,-.05){$=$} \put(-#2,0){\oval(#1,#3)[rb]}
  \put(-#2,-#3){\oval(#1,#3)[lt]}
  \put(-#1,-#3){\lower 1.4mm\hbox{\hspace*{-1.5mm}$\dis =#4$}}
  \end{picture}}}

\def\fall#1{\unitlength .7cm \begin{picture}(.6,.5)
   \put(.34,.2){\circle{.6}} \put(.2,.04){{\ft\bf #1}} 
   \end{picture}\,}

\documentclass[12pt]{article} \usepackage{amssymb}
\begin{document}

\vspace*{-12mm} 
\vskip 8mm

\rightline{ITP-UH-10/16}

\begin{center}
  {\large\bf Hamiltonian YM 2+1: 
             note on point splitting regularization } \\[6mm]
                Hermann Schulz 
              \footnote{ he.schulz38@web.de} \\[4mm]
{\small\sl
Institut f\"ur Theoretische Physik, Leibniz Universit\"at Hannover, \\
Appelstr.~2, D--30167 Hannover, Germany } \\[11mm] 

\end{center}

  \ct{\bf Abstract}

The Hamiltonian of 2+1 dimensional Yang Mills theory was
derived by Karabali, Kim and Nair by using point splitting
regularization. But in calculating e.g. the vacuum wave functional
this scheme was left in favour of arguments. Here we follow up
a conjecture of Leigh, Minic and Yelnikov of how this gap might be 
filled by including all positive powers of the regularization 
parameter ($\ep \to +0$). Admittedly, though we concentrate on the 
ground state in the large $N$ limit, only two such powers could be 
included due to the increasing complexity of the task. 

\vskip 1mm \hspace*{1cm} \parbox[t]{14cm}{
$\;\;\;\;\;${\bf Contents} \\[2mm]
1 \, The Problem \hspace{\fill} 1 \hspace*{3cm} \\
2 \, Closed expressions for the kernels $\o$ and $\O$ in $\cl T$
          \hspace{\fill} 3 \hspace*{3cm} \\
3 \, Expansion of $\o^{\heartsuit}\,$, $\o^{\diamondsuit}$ and 
     $\O$ in powers of $\ep$ \hspace{\fill} 5 \hspace*{3cm} \\
4 \, Coincidence limits \hspace{\fill} 8 \hspace*{3cm} \\[-1mm] 
    \hspace*{5.6mm} \rule[1mm]{6mm}{.1mm} \, {\ft\sl 
    end of the general analysis that keeps $\ep$ arbitrary} 
    \, \rule[1mm]{6mm}{.1mm} \\
5 \, Results of $\cl T$--application \hspace{\fill} 13 \hspace*{3cm} \\
6 \, The first steps in solving $\,\,\cl T \,\;\cl P \; = \;\cl E 
     - \cl V\,$ \hspace{\fill} 15 \hspace*{3cm} \\
7 \, Conclusions \hspace{\fill} 17 \hspace*{3cm} \\
$\;\;$  Appendices \hspace{\fill} 18 \hspace*{3cm} \\  
$\;\;$  References \hspace{\fill} 23 \hspace*{3cm} }

\let\dq=\thq \renewcommand{\theequation}{1.\dq}
                           \setcounter{equation}{0}
\section{The Problem}

This report is highly technical, treates a very special topic,
can not even reach the final answer to and is therefore not
considered for publication. Detailed Introductions are found in 
\cite{kkn}, \cite{kn}, \cite{lmy}, \cite{fks}, \cite{ich}. Let
us jump over them here. Some are lengthy (e.\,g.\ 40 pages 
in \cite{ich}). 

According to Karabali, Kim and Nair \cite{kkn} the Hamiltonian can
be prepared to act in the space of wave functionals depending on
only the currents $j^a$. It reads
\be{1.1}
    T + V = m \int\!\o_r^a \d_{j_r^a} + 
            m\int\!\!\int^{\prime}\!\O_{r r'}^{a b}
    \, \d_{j_r^a} \d_{j_{\rp}^b} + V \quad , \quad V = {N\0{m\pi}}\int
    \( {\ov \6} j_r^a \)\, {\ov \6 } j_r^a  \quad . \quad
\ee % 1.1
Here $a$ is the color index running from $1$ to $n \gll N^2-1 \,$,
$m = e^2 N / 2 \pi$ ($e$ the coupling), and $\int$ is shorthand for
$\int\!d^2 r$ running over the space of YM 2+1\,. Also $\int' = \int d^2r'\,$,
$\int'' = \int d^2r''$ etc\,. An Index $r$ on a quantity means that
it is a function of $\vc r = (x,y)\,$. Through $x-iy =z$, $x+iy = \ov z$
we have $r^2 = z\ov z$ and may define $\6 \gll d/dz$ and $\ov \6 \gll 
d/d\ov z\,$. By $\gll$ or $\glr$ the object near to the colon is defined. 

In strict temporal gauge there are only $2*n$ real gauge filds
$A^a_1\,,\;A^a_2\,$, combined to antihermitean and traceless
$N\times N$ fields by $A_j=-iA_j^aT^a$ or even to $A = (A_1+iA_2)/2\,$.
$T^a$ are the $n$ traceless generators of SU(N)\,: $\lk T^a, 
T^b \rk = i f^{abc} T^c \; , \, \Tr \( T^a T^b \) = {1\02} \d^{ab} 
\; , \; T^a T^a = n/(2N)\;$. Matrices $M \in\;$SL(N,C) 
parametrize the gauge fields by
\be{1.2}
  A = - (\6 M)M^{-1} \;\; , \;\; A = -iT^a A^a \;\; , \;\; A^a = 
  i\,2 \Tr \(T^a A \) \;\; , \;\; A^{ab} = -f^{abc} A^c \quad . \quad
\ee % 1.2
The above third relation rests on $2 \Tr \( T^a T^b \) = \d^{ab}\,$.
But it is also obtained by using $M^{ab} = 2\Tr\(T^a M T^b M^{-1}\)\,$
in $\lk (\6 M)M^{-1}\rki^{ab}\,$.
The single-indexed fields $A^a$ specify the functional derivatives 
in \eq{2.3} below.

The arguments of the functional derivatives in \eq{1.1} are
the currents
\be{1.3}
     j^a = 2 \Tr \,\( T^a j \) \;\, \hbox{with}\;\, 
     j = (\6 H)H^{-1} = T^a j^a \;\, , \;\,j^{ab} 
       = -if^{abc} j^c \;,\; j^a = {i\0N} f^{abc}j^{bc} \;, \quad
\ee % 1.3
of the WZW model, where $H = M^\dagger M\,$ is a gauge invariant. 
The last relation in \eq{1.3} derives via 
$ H^{ab} = 2 \Tr \( T^a H T^b H^{-1} \)\,$.
Again a superscript $r$ stands for $\vc r\,$ but mostly 
$H_r$ is considered as depending on $z$ and $\ov z\,$: 
$H_r = H_{z \ov z}\,$, $H_\rp = H_{z^\prime \ov z^\prime}$ etc. 
Round brackets are used to stop the action of 
a differentiation at the right bracket.
To distinguish fundamental from adjoint traces we shall write $\Tr$ for
the first and $( \; ... ¸\; )^{aa}$ for the latter. Under an adjoint
trace inner color--index--pairs are often omitted as e\,.g.\ in 
$\(j k \ell\)^{aa} \gll j^{ab} k^{bc} \ell^{ca}\,$. 

As long as the regularization parameter $\ep$ is kept non--zero
positive the functional Schr\"o\-din\-ger equation $(T+V)\,\psi\!\lk j_r^a \rk 
= E\,\psi\!\lk j_r^a \rk$ is fully regularized. There are no
singularities ({\sl \glqq no field theory\grqq } \anke apart from the
continuum of variables). At first the solution $\psi$ is to be obtained ($\psi$
depending on $\ep\,$, of course). The limit $\ep \to +0$
is allowed only afterwards. The expressions $\o_r^a$ and $\O_{r r'}^{a b}$
in \eq{1.1} can be booked down in closed form (sction 2)
{\sl and} can be fully expanded as power series in $\ep$ with all 
coefficients finite (section 3).

Former treatments perfomed $\ep \to +0$ too early, so they either
needed arguments or/and normal ordering or/and contact to known limiting
cases. Fine. Note that, thereby, all these authors accepted a certain 
break in the consequent working through the Karabali--Nair setup. The 
reason clearly is in the overwhelming complications otherwise. At least
Leigh, Minic and Yelnikov \cite{lmy} made an attempt in their appendix A
({\small \sl Regulated Computations}). They could not reach the expected
result. However they raised a conjecture about, what would have to 
be done. In the present report we follow it up and can justify their idea.
Admittedly the desired final result will be not reached here as well
\anke though by quite different reasons.  

Adopting the ansatz $\psi = e^P$ for the wave functional, with $H$ from
\eq{1.1} and with $e^{-P} \d_{j^a} e^P = P^{\prime a} +  \d_{j^a}\,$  
in mind $\,\( P^{\prime a} \gll \d_{j^a} P \)$ the Schr\"odinger equation
turns into
\be{1.4}
   m \int\!\o_r^a P_r^{\prime a} + m\int\!\!\int^{\prime}
   \!\O_{r r'}^{a b} \,\( \d_{j_r^a} P_{r'}^{\prime a}
 + P_r^{\prime a} P_{r'}^{\prime b} \)  = E - V  \quad . \quad
\ee % 1.4
Neither $\o$ nor $\O$ depend on $m\,$ (see section 2). By organizing
$P$ in powers of $1/m^2\,$, i.e. $P = P_0 +P_1 +P_2 +P_3 + \cdots$ with
$P_n \sim 1/m^{2n}\,$, \eq{1.4} decomposes in recursive equations \cite{kkn}
for $P_n\,$. In this note we concentrate on the leading nontrivial
term $P_1$ towards large $N$, i.e. $m \to \infty$ (the constant $P_0$
may be set equal to zero). For this task the term quadratic in $P$
in \eq{1.4} may be deleted since $\sim 1/m^4$. The now linear equation
for $P_1$ can simply be read off from \eq{1.4}. But in booking it down,
let us multiply the $P_1$--equation by $\pi m/N$ and rescale $P_1$, $E$ and $V$
to reach a more convenient form\,:
\be{1.5}
  \cl T \,\;\cl P \; = \;\cl E - \cl V \quad , \quad \hbox{where} \quad 
\ee % 1.5
\be{1.6}
   \cl T = {T \0 m} = \!\int\!\o_r^a \d_{j_r^a}
    + \!\int\!\!\!\int^{\prime}\!\O_{r r'}^{a b}
   \,\d_{j_r^a}\d_{j_{\rp}^b} \quad \hbox{and} \quad
\ee % 1.6
\be{1.7}
   \cl P = {\pi m^2 P_1 \0 N} \quad , \quad \cl E  = 
  {\pi m E \0 N} \quad , \quad \cl V = {\pi m V \0 N } 
   \!\int (\ov\6 j^a) (\ov\6 j^a) \quad . \quad
\ee % 1.7

All the experts \cite{kkn} to \cite{fks} agree, that the solution to
\eq{1.5} at $\ep \to +0$ is given by
\be{1.8}
   \cl P = -{1\02}\,\cl V = -{1\02}\int\! \({\ov \6} j_r^a \)\,{\ov \6 } j_r^a
    \quad \hbox{or} \quad P_1 = -{\pi \0 2 N m^2 }\int
    \( {\ov \6} J_r^a \)\, {\ov \6 } J_r^a \quad \quad
\ee % 1.8
where $P_1$ is rewritten in terms of the original currents 
$J^a= N j^a/\pi\,$ in \cite{kkn} and \cite{lmy} (but $j^a$ is $-J^a$
in \cite{fks}).

For solving $\,\cl T \,\cl P  = \cl E - \cl V\,$ we will first observe 
that $\cl T$ can be written as a power series in $\ep\,$.
If the unknown $\cl P$ is written as a power series aswell, a 
coefficient comparison yields to a set of equations, one for each 
given $\ep$--power. We emphasize that {\bf \,a\,l\,l\,} positive 
powers of $\ep$ are included in Secions 2 and 3. Also Section 4
does so \anke only the list \eq{4.10}, \eq{4.11} of invariants
remains too poor there.      

To learn about the solution step by step, one can break apart
at a given $\ep^n$\,: \glqq step $n$\grqq\ neglects powers
higher than $n\,$. Each step ends up with a system of equations
for the pefactors of the holomorphic invariants involved. Then
$\ep \to 0$ always leads to $\cl P = c_0 \, \cl V\,$. As will be 
seen in step 2 the result for $c_0\,$ varies under change 
of the highest included power $n\,$. Ultimately $n \to \infty$
is required \anke a strong support of the conjecture in \cite{lmy}. 

The solutions to step 1 and step 2 are obtained in the sequel.
Also step 3 is attacked but will remain incomplete. This is all we 
were able to really work through.

\let\dq=\thq \renewcommand{\theequation}{2.\dq}
                           \setcounter{equation}{0}
% Section 2
\section{Closed expressions for the kernels \boldmath$\o$ and \boldmath$\O$ in \eq{1.6}}

To derive (rederive in essence) these kernels we go a few steps
back in the Karabali--Kim--Nair analysis \cite{kkn}. Thereby special
attention is paid to maintain all $\ep$--dependence.
The regularised kinetic energy $\cl T := T/m$ originally reads
\be{2.1}
   \cl T = {\pi\0 N}\int\!\int^\prime\!\int^{\prime\prime} H_r^{ab}
  \; \ov{\cl G }^{au}_{r \rp} \; \ov{p}_{\rp}^u \; 
    {\cl G }^{bv}_{r \rpp} \; p_{\rpp}^v  \quad . \quad
\ee % 2.1 
Here the regularized Green functions are
\bea{2.2} 
 \ov {\cl G}^{au}_{r \rp} &=& \ov G_{r \rp} \( \d^{au} 
  - e^{ - (\vcsm r - \vcsm r^{\prime})^2 / \ep}
   \( H_{z \ov z'}\; {H_{z' \ov z'}}^{\! -1}
  \)^{au} \; \) \quad , \quad \nonu \\
 {\cl G}^{bv}_{r \rpp} &=& G_{r \rpp} \( \d^{bv} 
  - e^{ - (\vcsm r - \vcsm r^{\prime\prime})^2 / \ep} 
    \( {H_{z'' \ov z}}^{\! -1} \, H_{z'' \ov z''} \)^{bv} \; \) 
     \quad  , \quad  
\eea % 2.2
and the operators $\ov p$, $p$ are linear in functional differentiations
with respect to $A^{a*}$ and $A^a\,$, respectively\,:
\be{2.3}
    \ov{p}_{\rp}^u = - \ov \6'\,
             ( M_{\rp}^\dagger )^{ud} \d_{A^{d *}(\vcsm r')}  
    \quad , \quad
    \; p_{\rpp}^v = \6^{\prime\prime}
             ( {M_{\rpp}}^{\!-1} )^{vc} \d_{A^c(\vcsm r^{\prime\prime})}  
      \quad , \quad
\ee % 2.3 
see \eq{1.3}\,.
The prefactors $\ov G$ and $G$ in \eq{2.2} are the bare Green functions
\be{2.4}
     \ov G_{r \rp} = {1\0 \pi} {\ov z -\ov z' \0 (\vcsm r - 
                      {\vcsm r}^\prime)^2 +\varepsilon^2}
        \quad , \quad
       G_{r \rpp} = {1\0 \pi} {z -z^{\prime\prime} \0 (\vcsm r - 
                 {\vcsm r}^{\prime\prime})^2 +\varepsilon^2} \quad , \quad
\ee % 2.4
Since the singularity, which is controlled by $\varepsilon$ here, is
compensated by the round brackets in \eq{2.2} ($\ep$ positive. Note that
$\ep \neq \varepsilon$), one may perform $\varepsilon \to +0$ there. 
But otherwise it may happen that the limit $\varepsilon \to +0$ must wait 
at least until $\ov \6\,\ov G_{r \rp} = \d (\vc r -\vc r')$ or 
$\6\,G_{r \rpp} = \d (\vc r -\vc r^{\prime\prime})$ are 
reached.\footnote{Just to illustrate this let us ignore the 
        above regularization details for a moment. It is a digression 
        but one that could support confidence in the whole setup. 
        We replace the round brackets in \eq{2.2} by $\d^{au}$ and 
        $\d^{bc}$, respectively. Now the Delta functions just mentioned 
        come into play by partial integrations in \eq{2.1}. $H^{ab}= 
        M^{\dagger a g} M^{gb}$ combines with the $M$'s in \eq{2.3} 
        to $\d^{cd}$\, ($M$ does not depend on $A^\ast\,$), and things 
        turn into familar quantum mechanics\,:
        \vskip -4mm  \hspace*{6.1cm} $\dis
        \cl T \, = \, -{\pi\0 N} \int \d_{A^c *(\vcsm r)} 
        \d_{A^c(\vcsm r)} \, = \, {1\0 m} \; {e^2\0 2} \int {1\0 i} 
        \,\d_{A_j^a(\vcsm r)} {1\0 i} \d_{A_j^a(\vcsm r)} \;\; . $ \\[-4mm]  
        End of the digression.}

Since the kinetic energy $\cl T$ operates in the space of $\psi\!\lk j_r^a \rk\,$ 
it can be further reformulated. When applying \eq{2.1} 
\be{2.5}
  p_{\rpp}^v \psi = i\,\6^{\prime\prime} H_{\rpp}^{ev}\, \d_{j_{\rpp}^e}
  \psi \quad
\ee % 2.5
is needed first. Appendix A starts with a short derivation of \eq{2.5}\,.
Hence over this $\psi$--space \eq{2.1} becomes
\be{2.6}
   \cl T = {\pi\0 N}\int\!\int^\prime\!\int^{\prime\prime} H_r^{ab}
  \; \ov{\cl G }^{au}_{r \rp} \; \ov{p}_{\rp}^u \; 
    {\cl G }^{bv}_{r \rpp} \; 
    i \6^{\prime\prime} H_{\rpp}^{ev} \d_{j_{\rpp}^e} \quad . \quad
\ee % 2.6
$\ov{p}_{\rp}^u$ does {\sl\,n\,o\,t\,} commute with ${\cl G }^{bv}_{r
\rpp}\,$, see \eq{3.8} below. But note that even positive powers of 
$\ep$ are under study here. Hence by the chain rule the functional 
derivative in $\ov{p}_{\rp}^u\,$, \eq{2.3} splits $\cl T$ into three 
terms (acting only on ${\cl G}^{bv}\,$, only on $H^{ev}$ and only at 
the right end (i.e. on $ \d_{j_{\rpp}^e} \psi\,$), respectively.

The first two terms are linear in $\d$, the third term is quadratic.
Hence, with the notation of \eq{1.6}, there will be {\sl\,t\,w\,o\,} 
contributions to $\o_r^a\,$ \anke and hence in total three to $\cl T\,$:
\be{2.7}
  \cl T = \cl T_1 + \cl T_2 + \cl T_3 \quad , \quad
  \cl T_1 = \!\int\! \o_r^{\heartsuit a}\,\d_{j_r^a} \ \ , \ \
  \cl T_2 = \!\int\! \o_r^{\diamondsuit a}\,\d_{j_r^a} \ \ , \ \
  \cl T_3 = \!\int\!\!\!\int^{\prime}\!\O_{r r'}^{a b}
             \,\d_{j_r^a}\d_{j_{\rp}^b}     \quad . \quad
\ee % 2.7
$\cl T_1$ is studied with much detail in Appendix A since this
has possibly never been done. But $\cl T_2$, $\cl T_3$ merely 
recapitulate \cite{kkn} in Appendix A. Here we summarize the results\,:
\bea{2.8}
    \o_r^{\heartsuit a} &=& -{i\0 N}\,\lk \6 -j_r \rki^{ab} f^{buv} \!\int^\prime\! 
   { \,e^{ \,- 2\, (\vcsm r -{\vcsm r}^{\,\prime} )^2 / \ep } \0
     \pi (\vcsm r -{\vcsm r}^{\,\prime})^2 }
    \,\( H_{z \ov z'}\; {H_{z' \ov z'}}^{\!\! -1} 
     H_{z' \ov z}\; {H_{z \ov z}}^{\!\! -1} \)^{uv} \quad , \quad 
 \\ \label{2.9}
    \o_r^{\diamondsuit a} &=& {i\0 N}\,f^{auv}\,\Theta_{rr}^{uv} \quad , \quad 
 \\ \label{2.10}
    \O_{r \rp}^{ab} &=& {1\0 N}\,\lk\6 -j_r\rki^{ac}\;\Theta_{r \rp}^{cb} 
    \qquad \hbox{with} \quad 
 \\ \label{2.11}
    & & \Theta_{r \rp}^{ab} \,=\, -\pi \!\int^{\prime\prime} \!
    \ov {\cl G }^{ua}_{\rpp r}\;  H_{\rpp}^{uv} \;
    \( \6^\prime {\cl G }^{vc}_{\rpp \rp} \)  \( H_{\rp}^{-1}\)^{cb} \quad . \quad
\eea % 2.8 bis 2.11     
By $\Theta = \pi \L\,$ the notation of \cite{kkn} is reached. There we are. And now\,?

\let\dq=\thq \renewcommand{\theequation}{3.\dq}
                       \setcounter{equation}{0}
% Section 3
\section{Expansion of \boldmath$\o^{\heartsuit}$, \boldmath$\o^{\diamondsuit}$
            and \boldmath$\O$ in powers of \boldmath$\ep$}
\vskip -3mm
\subsection{ \boldmath$\o^{\heartsuit}$}   % \S 3.1

This is the sometimes omitted term of order $O(\ep)\,$. It is tempting
to perform the shift of variables $\vc r' \to \vc r' +\vc r$ under
the primed integral in \eq{2.8} and then writing the shifts by
$\ov z$ or $z$ in the $H$--arguments as Taylor expansions\,:
\be{3.1}
   \o_r^{\heartsuit a} = -{i\0 N}\,\lk \6 -j_r \rki^{ab} f^{buv} H_r^{vw}\!\int^\prime\! 
   { \,e^{ \,- 2\, r^{\,\prime\,2} / \ep } \0
     \pi r^{\,\prime \,2 } } \sum_{s,t=0}^\infty {z^{\prime\,s}\,
     {\ov z}^{\prime \,t} \0 s!\, t!} \,\hat{\6}^s \,\underline{\ov \6}^t
    \,\( \underline{H}_{z \ov z}\; {\hat{\underline{H}}_{z \ov z}}^{\!\! -1} 
     \hat{H}_{z \ov z}\; \)^{uw} \quad . \quad 
\ee % 3.1 
Here $\hat \6$ acts on $\hat H$ only and $\underline{\ov \6}$ only on
$\underline H\,$. The primed integral is nonzero only for $s=t$, and
at $s=t=0$ \eq{3.1} vanishes due to $(H H^{-1} H H^{-1})^{vu} = \d^{vu}\,$.
Now the integration can be done. Using $\int e^{-2 r^2/\ep } (r^2)^{s-1}/\pi
= (\ep/2)^s (s-1)!$ we get
\be{3.2}
   \o_r^{\heartsuit a} = -{i\0 N}\,\lk \6 -j_r \rki^{ab} f^{buv} H_r^{vw}
   \sum_{s=1}^\infty { (\ep/2)^s \0 s\; s!} \hat{\6}^s \underline{\ov \6}^s
    \,\( \underline{H}_{z \ov z}\; {\hat{\underline{H}}_{z \ov z}}^{\!\! -1} 
     \hat{H}_{z \ov z}\; \)^{uw} \quad . \quad 
\ee % 3.2
$\hat\6$ acts on two of the $H$'s. Correspondingly, $\hat\6^s$ can be decomposed 
binomically by
\be{3.3}
 \hat\6^s = \sum_{k=0}^s \, { s! \0 k! \, (s-k)! } \;\6_{\hbox{\ft first}}^{k}
        \; \6_{\hbox{\ft second}}^{s-k} \qquad , \quad \hbox{hence} \qquad
\ee % 3.3
\vskip -8mm
\bea{3.4}
   \sum_{s=1}^\infty \cdots \;\glo{8}\, \sum_{s=1}^\infty \sum_{k=0}^s
    { (\ep/2)^s \0 s\; k!\,(s-k)! }
    \lk \hbox{\Large $($}\, {\ov\6}^s H ( \6^k H^{-1} )\,\hbox{\Large $)$}\, 
     \6^{s-k} H \rki^{uw} \quad \\ \label{3.5}
     \glu{8}\, \sum_{s=1}^\infty \sum_{k=0}^s
    { (\ep/2)^s \0 s\; k!\,(s-k)! } \lk \( \ov \6^s B(k) \) A(s-k) H
    \rki^{uw}  \quad  \quad
\eea % 3.4 , 3.5
with the definitions
\be{3.6}
 B(p) \gll H \6^p H^{-1} \quad \hbox{and} \quad A(p) \gll \( \6^p H\) H^{-1}
 \quad , \quad B(p)^{ab} = A(p)^{ba} \quad , \quad
\ee % 3.6
since $(H^{-1})^{ab} = H^{ba}$ from $H^{ab}=2\,\Tr \(T^a H T^b H^{-1}\)\,$.	
Inserting \eq{3.5} into \eq{3.2}, redefining $k \to s-k$ and using
$B^{ab} = A^{ba}$ one obtains 
\be{3.7}
  \o^{\heartsuit a} = {i\0 N}\,\lk \6 -j \rki^{ab} f^{buv} 
   \sum_{s=1}^\infty \sum_{k=0}^s { (\ep/2)^s \0 s\; k!\,(s-k)! } 
  \( B(k)\, \ov \6^s A(s-k) \)^{uv} 
  \quad . \quad 
\ee % 3.7 
For the final $\o^{\heartsuit a}$--version
\be{3.8}
  \o^{\heartsuit a} = {i\0 N}\,\lk \6 -j \rki^{ab} f^{buv} \sum_{s=1}^\infty \sum_{k=0}^s
    { (\ep/2)^s \0 s\; k!\,(s-k)! } \; M(k,s,s-k,0)^{uv} \qquad
\ee % 3.8
the more general definition
\be{3.9}
       M(a,b,c,d) \gll B(a) \,\ov\6^b A(c) B(d) 
\ee % 3.9
has been introduced. Note that $B(0)=1$. With $j = (\6 H) H^{-1}$ it follows
from \eq{3.6} that $B(p) =  \lk \6-j\rki^p\,1\,$ and in particular {\small
\bea{3.10}
  B(1)\!\!&=&\!\! -j \;\; , \;\; B(2) = -\6 j + j\,j \;\; , \;\; 
    B(3) = -\6^2 j +(\6 j)\,j +2 j\,\6j - j\,j\,j 
    \quad \hbox{and} \qquad \qquad \nonu \\
  B(4)\!\!&=&\!\! -\6^3 j  +(\6^2 j)\,j +3\,j\,\6^2 j +3\,(\6 j)\,\6 j 
        -(\6 j)\,j\,j -2\,j\,(\6 j)\,j -3\,j\,j\,\6 j +j\,j\,j\,j 
          \;\; . \qquad   
\eea % 3.10
} \noindent
$\!\!$For special $A$'s use $A^{ab}=B^{ba}\,$ ($A(0)=1,\, A(1)=j,\, \cdots$). 
\eq{3.8} shows that the contribution $\o^{\heartsuit a}$ to $\o^a$ 
would have been bypassed by a too rush $\ep \to +0\,$.

\subsection{ \boldmath$\Theta$} % \S 3.2

\eq{2.9} and \eq{2.10} show that both quantities, $\o^{\diamondsuit}$
and $\O\,$, are traced back to the object $\Theta_{r \rp}^{ab}\,$.
The rhs of \eq{2.11} has not quite the form of {\small\sl something}$^{ab}$.
But the indices on the troublemaker $\ov {\cl G}_{\rpp r}^{ua} \,$
(see \eq{2.2}\,) are easily reversed by
$ \( H_{z^{\prime\prime} \ov z}\; H_r^{-1}\)^{au} = 
   H_{z^{\prime\prime} \ov z}^{ac}\, H_r^{uc} = 
  \( H_r\; H_{z^{\prime\prime} \ov z}^{-1}\)^{ua}\,$.
In the sequel let us omit (but keep in mind) the superscipts $^{ab}$ on both sides.
Also let $e_{\rpp r}$ stand for $\exp\(-(r^{\prime\prime} - \vc r)^2/\ep\)\,$.
In otherwise full detail \eq{2.11} now reads
\be{3.11}
  \Theta_{r \rp} = -\pi \!\int^{\prime\prime} \!
    \ov G_{\rpp r} \( 1 - e_{\rpp r} H_r 
          H_{z^{\prime\prime} \ov z}^{-1} \) H_\rpp 
   \hbox{\Large $($}\,\6^\prime G_{\rpp \rp} \( 1 - e_{\rpp \rp} 
   H_{z^\prime \ov z^{\prime\prime}}^{-1} H_\rp \) \,\hbox{\Large $)\,$}
   H_{\rp}^{-1} \quad . \quad
\ee % 3.11
Now $\ov G$ and $G$ may be replaced by their bare versions $(\e=0)$. 
Note that $\6^\prime $ and $G_{\rpp \rp}$ can be commuted since 
the Delta function would be multiplied by the vanishing round bracket. 
Hence                        
\bea{3.12} 
  \Theta_{r \rp} &=& {1\0 \pi} \!\int^{\prime\prime} \!
  { e_{\rpp \rp} \0 ( z^{\prime\prime} -z)
                    ( \ov z^{\prime\prime} -\ov z^\prime )}
   \Bigg( {\ov z^{\prime\prime} -\ov z^\prime \0 \ep} 
              H_\rpp H_{z^\prime \ov z^{\prime\prime}}^{-1} 
   + H_\rpp \( \6^\prime H_{z^\prime \ov z^{\prime\prime}}^{-1} H_\rp \)
     H_{\rp}^{-1}  \hspace*{2cm} \nonu \\
   & & \hspace*{1mm}
   - e_{\rpp r} {\ov z^{\prime\prime} -\ov z^\prime \0 \ep}
       H_r H_{z^{\prime\prime} \ov z}^{-1} 
       H_\rpp H_{z^\prime \ov z^{\prime\prime}}^{-1}
   - e_{\rpp r}  H_r H_{z^{\prime\prime} \ov z}^{-1} 
      H_\rpp \( \6^\prime H_{z^\prime \ov z^{\prime\prime}}^{-1} H_\rp \) 
      H_{\rp}^{-1} \Bigg) \quad . \quad 
\eea % 3.12
\eq{3.12} is in perfect agreement with the four terms I to IV in 
\cite{kkn} (the $\L$ there is $=\Theta/\pi\,$)\,. In \cite{kkn}  
I to IV are eqs. (4.7 b to e). (Some of their subsequent equations 
are not free from errors or typos). What follows here is new ground.

Towards evaluation of the integral the shift $\vc r^{\prime\prime} \to
\vc r^{\prime\prime} + \vc r^\prime$ is a first useful step\,:
\bea{3.13}
   & & \hspace*{-17mm}
   e_{\rpp \rp} \to \exp{\( - r^{\prime\prime \,2} / \ep \,\)} 
    \; , \;
   e_{\rpp r} \to \exp{\(- r^{\prime\prime \,2} / \ep  
       - \s \ov z^{\prime\prime} / \ep 
       - \ov\s z^{\prime\prime} / \ep  
       - \s \ov\s / \ep \,\)} \nonu \\[1mm]
   & & \hspace*{-17mm} \hbox{where}
    \quad \s \gll z^\prime -z \quad \hbox{and} \quad 
    \ov \s \gll \ov z^\prime -\ov z \quad . \qquad      
\eea % 3.13
\vskip -11mm
\bea{3.14}
  \Theta_{r \rp} \!\!&=&\!\! {1\0 \pi} \!\int^{\prime\prime} \! \Bigg( \;
  { e^{- r^{\prime\prime \,2} / \ep}
    \0 \ep ( z^{\prime\prime} + \s ) }
    \,H_{z^\prime + z^{\prime\prime}\;\ov z^\prime + \ov z^{\prime\prime}}
    \, H_{z^\prime \; \ov z^\prime + \ov z^{\prime\prime}}^{-1}  
       \hspace*{7cm} \nonu \\ 
  & & \hspace*{11mm} 
     + { e^{- r^{\prime\prime \,2} / \ep}
    \0 \ov z^{\prime\prime} ( z^{\prime\prime} + \s ) }
    \,H_{z^\prime + z^{\prime\prime}\;\ov z^\prime + \ov z^{\prime\prime}}
    \( \6^\prime\, H_{z^\prime \; \ov z^\prime + \ov z^{\prime\prime}}^{-1}
      H_\rp \) H_{\rp}^{-1} \nonu \\    
  & & \hspace*{11mm} 
    - { e^{- 2 \, r^{\prime\prime \,2} / \ep}
    \0 \ep ( z^{\prime\prime} + \s ) }
       \,e^{- \s \ov z^{\prime\prime} / \ep} 
       \,e^{- \ov \s z^{\prime\prime} / \ep} \,e^{- \s \ov\s / \ep}
       \,H_r H_{z^\prime + z^{\prime\prime}\; \ov z}^{-1}
    \,H_{z^\prime + z^{\prime\prime}\;\ov z^\prime + \ov z^{\prime\prime}}
    \, H_{z^\prime \; \ov z^\prime + \ov z^{\prime\prime}}^{-1} \nonu \\ 
  & &  \hspace*{-16mm} 
    - { e^{- 2 \, r^{\prime\prime \,2} / \ep}
    \0 \ov z^{\prime\prime} ( z^{\prime\prime} + \s ) }
       \,e^{- \s \ov z^{\prime\prime} / \ep} 
       \,e^{- \ov \s z^{\prime\prime} / \ep} \,e^{- \s \ov \s / \ep}
       \,H_r H_{z^\prime + z^{\prime\prime}\; \ov z}^{-1}
    \,H_{z^\prime + z^{\prime\prime}\;\ov z^\prime + \ov z^{\prime\prime}}
    \( \6^\prime\, H_{z^\prime \; \ov z^\prime + \ov z^{\prime\prime}}^{-1}
      H_\rp \) H_{\rp}^{-1} \; \Bigg) \qquad
\eea % 3.14
now shows that all the $z^{\prime\prime}$ and $\ov z^{\prime\prime}$
in the $H$--arguments can be transformed into powers by means of Taylor 
expansions. Hence all the integrals to be done will be of the form
\be{3.15}
  I_{p,q}(\ep,\s ) = {1\0 \pi} \!\int^{\prime\prime}\!
  { e^{- r^{\prime\prime \,2} / \ep} \0 z^{\prime\prime} +\s }
  \; z^{\prime\prime \, p} \; \ov z^{\prime\prime \; q-1}
  \quad . \quad
\ee % 3.15
The four lines of \eq{3.14} become
\bea{3.16} \!\!
  \Theta_{r \rp} \!\!&=&\!\! {1\0 \ep} \sum_{p=0}^\infty \sum_{q=1}^\infty
   \,{1\0 p!\,(q-1)!}\, I_{p,q}(\ep,\s) \; \ov \6^{\prime \; q-1} \6^{\prime \; p}
   H_{\rp} H_{\rp}^{-1}  \nonu
                            \\[-6mm] & & \hbox{\tiny
                            \hspace{5.1cm} $\uparrow$
                            \hspace{6.1mm} $\bullet$
                            \hspace{5mm} $\bullet\!\uparrow$ 
                            \hspace{2.4mm} $\uparrow$} \nonu \\[2mm] 
   & &  
   + \sum_{p=0}^\infty \sum_{q=0}^\infty \, 
     \,{1\0 p!\,q!}\, I_{p,q}(\ep,\s ) \; \ov \6^{\prime \; q} \6^{\prime \; p}
     H_\rp \( \6^\prime H_{\rp}^{-1} H_\rp \) H_{\rp}^{-1} \nonu 
                            \\[-6mm] & & \hbox{\tiny
                            \hspace{4.3cm} $\uparrow$
                            \hspace{2.9mm} $\bullet$
                            \hspace{5.1mm} $\bullet\!\uparrow$ 
                            \hspace{8mm} $\uparrow$} \nonu \\[2mm] 
   & & \hspace*{-23mm}
       -{1\0 \ep} e^{-\s \ov \s /\ep} H_r \sum_{p=0}^\infty \sum_{q=1}^\infty
        \,{1\0 p!\,(q-1)!}\, I_{p,q}({\ep \0 2},\s) \ 
        \!\lk \ov \6^\prime -{\s\0\ep}\rki^{q-1} \!\lk \6^\prime 
            -{\ov \s \0 \ep} \rki^p
         H_{z^\prime \; \ov z}^{-1} H_\rp H_{\rp}^{-1}  \nonu  
                            \\[-6mm] & & \hbox{\tiny
                            \hspace{5.4cm} $\uparrow$
                            \hspace{20.3mm} $\bullet$
                            \hspace{16.6mm} $\bullet$
                            \hspace{4.5mm} $\bullet\!\uparrow$ 
                            \hspace{2.2mm} $\uparrow$} \nonu \\[2mm] 
   & &  \hspace*{-23mm}
    -e^{-\s \ov \s /\ep} H_r \sum_{p=0}^\infty \sum_{q=0}^\infty  
     {1\0 p!\,q!}\, I_{p,q}({\ep \0 2},\s)\, 
     \!\lk \ov \6^\prime -{\s\0\ep}\rki^q \!\!\lk \6^\prime 
         -{\ov \s \0 \ep}\rki^p
     \!H_{z^\prime \; \ov z}^{-1} H_\rp \( \6^\prime H_{\rp}^{-1}
     H_\rp\) H_{\rp}^{-1} \; .  \\[-6mm] & & \hbox{\tiny
                            \hspace{4cm} $\uparrow$
                            \hspace{15.1mm} $\bullet$
                            \hspace{16.1mm} $\bullet$
                            \hspace{4.8mm} $\bullet\!\uparrow$ 
                            \hspace{8.1mm} $\uparrow$} \nonu  
\eea % 3.16
The meaning of the subscipts\,: a dotted $\6^\prime$ acts {\sl \,o\,n\,l\,y\,}
on dotted $H$--arguments and an arrowed only on arrowed. In particular, 
$\ov \6^\prime$ in the last two lines does {\sl \,n\,o\,t\,} act on 
$\ov \s $ in the second square bracket. 

The integral $I\,$, \eq{3.15}, is evaluated in Appendix B
to be
\be{3.17}
  I_{p,q}(\ep,\s ) \; = \; (-\s )^{p-q} \!\int_{\s \ov \s }^\infty\!\! dt\;\, 
  t^{q-1} e^{-t/\ep}\,\;-\,\;\theta_{q>p}\; (q-1)! \,(-\s)^{p-q} \,\ep^q 
 \quad . \;\;  
\ee % 3.17
Here $\theta_{q>p}$ is 1 for $q>p$ and zero otherwise. Surprizingly,
the left half of \eq{3.17} does not contribute to \eq{3.16} at all.
The check to this nice outcome is left to the 
reader\footnote{\,Perform each of the four $p$--sums. Note that the 
  operator $\exp(-\s \6^\prime)$\hspace{-4.4mm}\lower 1.3mm\hbox{\tiny $\bullet$}
  \,\,\,\,changes a dotted $H$--index $z^\prime$ into 
  $z^\prime -(z-z^\prime) = z\,$. Show that the third line of \eq{3.16}   
  is the negative of the first one. \glqq But $I(\ep/2,\ldots)$ contains
  the wrong $\exp(-2 t/\ep)$\,!?\grqq . Well, use the $q$--sum
  to produce the missing $\exp(+t/\ep)\,$: \\ 
  $ \sum_{q=0}^\infty {1\0q!} \lk -t {\ov\6}^\prime/\s + t/\ep \rki^q  
  = {\rm e}^{\lk -t{\ov\6}^\prime / \s +t/\ep \rk} =
  \sum_{q=0}^\infty {1\0q!} \lk -t{\ov\6}^\prime/\s \rki^q \,
  {\rm e}^{+t/\ep}\,$. Show fourth line $= -$\,second.}.
Hence for use in \eq{3.16} the second term of \eq{3.17} suffices.  

Due to $\theta_{q>p}$ all $q$--sums in \eq{3.16} become
$\sum_{q=p+1}^\infty\,$. The factor $(q-1)!$ either compensates with 
the denominator or leaves a $q$ there. For just one more detail note 
that the ends of the second and fourth line may be written as \\[-3mm]
\be{3.18}
  \( \6^\prime H_{\rp}^{-1} H_\rp \) H_{\rp}^{-1} \;=\; 
       H_{\rp}^{-1}\,\( -j_\rp + j_\rp \) \quad . \quad  
       \\ \lower 3mm\hbox{\tiny
       \hspace{-6.4cm} $\uparrow$
       \hspace{2.85cm} $\uparrow$ 
       \hspace{9.5mm} $\uparrow$ } 
\ee % 3.18
For the further lengthy analysis we denote the four lines of 
\eq{3.16} by \fall 1 , \fall 2 , \fall 3 , \fall 4 . One obtains
\bea{3.19}
   \fall 1 + \fall 2 = {1\0\s} + {\cl C} j_{r^\prime} 
   + \6^\prime {\cl C} &\hbox{with}&
   {\cl C} = - \sum_{p=1}^\infty \sum_{q=p+1}^\infty
  {(-\s)^{p-q} \0 p!\, q } \e^q \ov \6^{\prime \; q} \6^{\prime \;p}
   H_{\rp} {H_{\rp}}^{\!\!-1} \; . \qquad 
                            \\[-6mm] & & \hbox{\tiny
                            \hspace{4.7cm} $\uparrow$
                            \hspace{3mm} $\bullet$
                            \hspace{4mm} $\bullet\!\uparrow$
                            \hspace{2mm} $\uparrow$} \nonu 
\eea % 3.19
More laboriously one arrives at
\bea{3.20}
   \fall 3 + \fall 4 = - {1\0\s} e^{-\s \ov \s /\ep} H_r\, 
        H_{z^\prime \ov z}^{-1} + {\cl E} j_{r^\prime} 
    + \6^\prime {\cl E} &\hbox{with}& \hbox{\hspace*{5.6cm}} \nonu \\
     & & \hspace*{-7cm}  
    {\cl E} =  \,e^{-\s \ov \s /\ep} H_r 
    \sum_{p=1}^\infty \sum_{q=p+1}^\infty 
    \( {\ep\0 2}\)^q
      \lk \ov \6^\prime -{\s\0 \ep} \rki^q 
      \lk\6^\prime -{\ov \s \0 \ep} \rki^p
     H_{z^\prime \ov z}^{-1} \, H_\rp H_{\rp}^{-1}  \;\, . \quad
                            \\[-5.2mm] & & \hspace*{-2.4cm} 
                            \hbox{\tiny
                            \hspace{9mm} $\uparrow$
                            \hspace{1.8cm} $\bullet$
                            \hspace{16.5mm} $\bullet$
                            \hspace{4.1mm} $\bullet\!\uparrow$
                            \hspace{2.3mm} $\uparrow$} \nonu
\eea % 3.20
Through \fall 1 + \fall 2 + \fall 3 + \fall 4 and with 
$ {\cl A}:={\cl C} + {\cl E}$ we obtain
\be{3.21}
  \Theta_{r \rp} = {1\0 \s } \( 1 - e^{-\s \ov \s /\ep} H_r 
     H_{z^\prime \ov z}^{-1} \) \,  + \, \6^\prime 
     \cl A_{r \rp} + \cl A_{r \rp}\,j_\rp \quad . \quad 
\ee % 3.21
\vskip -2mm
The double sum over $p$ and $q$ is common to ${\cl C}$ and
${\cl E}$, hence also to ${\cl A}\,$.
A rearrangement of this double sum comes into mind by
\be{3.22}
  \sum_{p=0}^\infty \sum_{q=p+1}^\infty =
  \sum_{p=0}^\infty \( \sum_{q=p+1}^\infty - \sum_{q=1}^\infty \)
   + DS = - \sum_{p=1}^\infty \sum_{q=1}^p + DS
   \quad \hbox{with} \quad DS := \sum_{p=0}^\infty \sum_{q=1}^\infty 
      \; . \quad
\ee  % 3.22
${\cl A}$ can be split into the part with
the finite $q$--sum and the part with $DS$. Let the latter 
part be called ${\cl F}$. Then
\bea{3.23}
 {\cl F} &=& \sum_{q=1}^\infty {1\0 q} \lk 
      - \(-{\e\0\s} \ov\6^\prime  \)^q + \(-{\e\0 2\s} 
          \lk \ov\6^\prime - {1\02} \rki^q \) \rk
         H_{z \ov z^\prime}\,H_{r^\prime}^{-1} \nonu 
                          \\[-5.6mm] & & \hspace*{1.8cm} 
                          \hbox{\tiny
                          \hspace{9mm} $\uparrow$
                          \hspace{2.6cm} $\uparrow$
                          \hspace{2.4cm} $\uparrow$} \nonu \\[2mm]
     &=& \lk \ln \( 1 + {\e\0\s} \ov \6^\prime \)
       -\ln \( 1 + {\e\0 2\s} \ov\6^\prime -{1\02}\) \rk
            H_{z \ov z^\prime}\,H_{r^\prime}^{-1} 
     = \ln (2)\, H_{z \ov z^\prime}\,H_{r^\prime}^{-1} 
                            \\[-4.2mm] & & \hspace*{1cm} 
                            \hbox{\tiny
                            \hspace{1.05cm} $\uparrow$
                            \hspace{2.8cm} $\uparrow$
                            \hspace{2cm} $\uparrow$} \nonu
\eea % 3.23
and this drops out in \eq{3.21} because
$ \6^\prime {\cl F} + {\cl F} j_{r^\prime} =0\,$.

Hence the suitable completion of \eq{3.21} is
\bea{3.24}
  \cl A_{r \rp} &=& \sum_{p=1}^\infty \,\sum_{q=1}^p 
  \,{1\0 p!\,q} (-\s)^{p-q} \,\Bigg( \,\ep^q \,\ov \6^{\prime \,q}  
         \(\6^{\prime \,p} H_\rp \) H_{\rp}^{-1}
        \hspace*{5cm} \nonu 
       \\[-6mm] & & \hbox{\tiny
       \hspace{4.1cm} $\uparrow$
       \hspace{4.6mm} $\bullet$ 
       \hspace{4.6mm} $\bullet\!\uparrow$ 
       \hspace{4.8mm} $\uparrow$ } \nonu \\[2mm]
     & & \hspace*{1.8cm} 
      - \,e^{ -\s \ov \s /\ep} H_r \( {\ep\0 2}\)^q 
       \lk \ov \6^\prime -{\s\0 \ep} \rki^q \lk\6^\prime -{\ov \s \0 \ep} \rki^p 
      H_{z^\prime \ov z}^{-1} H_\rp H_{\rp}^{-1} \Bigg) \quad . \quad
                            \\[-4.2mm] & & \hbox{\tiny
                            \hspace{5.2cm} $\uparrow$
                            \hspace{17.1mm} $\bullet$
                            \hspace{17.1mm} $\bullet$
                            \hspace{4.06mm} $\bullet\!\uparrow$ 
                            \hspace{2.26mm} $\uparrow$} \nonu  
\eea % 3.24

Two agreeable properties of \eq{3.19} might be emphasized.
All dependences on $\ep$ are contained (all positive and negative 
$\ep$ powers \anke imagine the exponential functions be expanded).
Secondly, there are no negative powers of $\s$ or $\ov \s$ in 
$\Theta_{r \rp}\,$. This is a welcome fact towards coincidence limits 
$\vc r^\prime \to \vc r$ which make $\s\,,\;\ov \s$ vanish. Such limits 
occur when two functional derivatives \anke remember 
$\cl T_3 = \!\int\!\!\!\int^{\prime}\!\O_{r r'}^{ab} \,\d_{j_r^a}
\d_{j_{\rp}^b}\,$ and $\,\O_{r \rp}^{ab} = {1\0 N}\,\lk\6 -j_r\rki^{ac}
\Theta_{r \rp}^{cb}\,$ \anke are applied to holomorphic invariants. 
$\,\Theta_{r \rp}$ is further processed in the subsections 4.2 and 4.3\,.

\let\dq=\thq \renewcommand{\theequation}{4.\dq}
                           \setcounter{equation}{0}

\section{ Coincidence limits}
\nopagebreak
\vskip -3mm
\subsection{ Holomorphic invariants} 

The mapping of $A$ to the $M$--space is not unique since
$A_r = - (\6 M_r) M_r^{-1}$ remains unchanged under 
\be{4.1}
  M\,\to\,M h^{\dagger}(\ov z)  \quad  
  \folg \;\; M_r^\dagger\,\to\, h(z) M_r^\dagger\, \;\; \hbox{and} \;\;
  H=M^{\dagger}M\,\to\, h(z) H h^{\dagger}(\ov z) \quad . \quad  
\ee % 4.1
Physics must not depend on the choice 
of $h\,$. The solution $\cl P$ to $\,\cl T \,\cl P  = \cl E - \cl V\,$
is \glqq physics\grqq . So $\cl P$ has to be holomorphic invariant.
$\cl P$ will turn out to be a linear combination of the invariants
listed in \eq{4.9} to \eq{4.11} below. By means of \eq{4.1}
one derives
\be{4.2}
  j = (\6 H) H^{-1} \,\to\, (\6 h) h^{-1} + h j h^{-1} \qquad \hbox{but}
 \qquad \ov\6 j \,\to\, h (\ov\6 j) h^{-1} \quad . \quad 
\ee % 4.2 
For convenience we might leave the above $N\times N$--matrix--language
and ask for the holomorphic transformation of $j^a$ and $j^{ab}$.
Let us also shorten the notation a bit by
\be{4.3}
   \ov\6 j \glr \ov\jmath \quad , \quad 
   \ov\6 j^a \glr \ov\jmath^a \quad , \quad 
   \ov\6 j^{ab} \glr \ov\jmath^{ab} \quad , \quad 
   \ov\6^2 j \glr \ov{\ov\jmath} \quad , \quad \hbox{etc.} \quad 
\ee % 4.3
The last equation in \eq{4.2} now reads 
$\ov\jmath \,\to\, h\, \ov\jmath\, h^{-1}$.
From \eq{1.2}, which now reads $\ov\jmath^a = 
2\,\Sp \( T^a \ov\jmath \)\,$, one derives that 
\be{4.4}
    \ov\jmath^a \, \to\, 2\,\Sp \( T^a\, h\, \ov\jmath\, h^{-1} \) 
   =  2\Sp \(h^{-1}\, T^a\, h\, T^b\)\,2\Sp \(T^b\, \ov\jmath \) 
   = h^{ab}\, \ov\jmath^b \quad , \quad 
\ee % 4.4
where $\, h^{ab} = 2\,\Sp\( T^a h T^b h^{-1} \) = 2\,\Sp\( T^b h^{-1} T^a h \)
         = (h^{-1})^{ba}\,$. 
Hence $ \ov\jmath^a\, \ov\jmath^a$ is invariant, and so is $\cl V =\int
\ov\jmath^a\,\ov\jmath^a\,$. From $j^{ab} = \( (\6 H)H^{-1} \)^{ab}$
and \eq{4.2} one obtains
\bea{4.5}
  j^{ab} &\to&  \( (\6 h H h^\dagger ) h^{\dagger \, -1} H^{-1} 
         h^{-1} \)^{ab} \, = \, \( (\6 h) h^{-1} + h j h^{-1} \)^{ab}
          \quad \hbox{but} \quad \nonu \\
  \lk\6-j\rki^{ab} &\to& \(\6 - (\6 h) h^{-1} - h j h^{-1} \)^{ab}
        \,=\, \( h \, \lk\6-j\rk \,h^{-1} \)^{ab} \quad . \quad
\eea % 4.5
The above details allow to recognize the following set
\be{4.6}
   Q_n = \int \ov\jmath^a \( \lk [\6-j]\,\ov\6 \rki^n \)^{ab}
          \ov\jmath^b \quad , \quad n = 0,1,2, \cdots\, \quad , \quad
\ee % 4.6  
as holomorphic invariants. Note that $Q_0 = \cl V\,$. 

The invariants are conveniently written as adjoint traces. \eq{4.9} 
below shows this trace for $Q_n\,$. It contains the operator
\be{4.7}
 \cl D \gll \6 - \lk j\; , \,\;\; \rk \;\quad {\rm where} \;\quad
      \lk j\; , \,\;\; \rk any \,\gll\, j\,any -any\,j \quad . \quad
\ee % 4.7 
Starting from \eq{4.6} the version
\eq{4.9} can be derived.\footnote{\, Consider $Q_n = 
     \int \ov\jmath^a \( \lk [\6-j]\,\ov\6 \rki^{n-1}\)^{ac}
     \,[\6-j]^{cb}\;\ov{\ov\jmath}{\,}^b\,$ and insert 
     $\,j^{cb} =-i\,j^d f^{dcb}\,$ and $\,\ov{\ov\jmath}{\,}^b 
     = i\,f^{buv}\;\ov{\ov\jmath}{\,}^{uv}/N \,$ at the right end.
     Now the Jacobi identity $\,f^{dcb}f^{buv} = -f^{udb}f^{bcv}
     -f^{cub}f^{bdv}\,$ leads to
     $\,[ \;\quad\; ]^{cb}\;\ov{\ov\jmath}{\,}^b\, =
     \,i\, f^{cbv} \lk \cl D \,\ov \6 \,\ov \jmath \rki^{bv} /N \,$
     \anke and so on.}

As is seen in section 5 (and latest in section 6) {\sl \,m\,o\,r\,e\,}
holomorphic invariants are to be included. They will all be of the form
\be{4.8}
      {1\0 N} \int \big(\, any \,\big)^{aa} \qquad \hbox{with}
      \quad any \to h\,any\,h^{-1} \quad . \quad   
\ee % 4.8    
Our attempts to solve $\,\cl T \,\cl P  = \cl E - \cl V\,$ (up
to some maximal $\ep$--power) involve the followoing invariants
\bea{4.9}
   Q_n &=& {1\0 N} \int \( \ov\jmath \lk\! \cl D \,\ov\6 \rki^n 
     \ov\jmath \)^{aa}  \quad , \quad \\  \label{4.10}
  R_{20} &=& {1\0 N} \int \big( \,\ov\jmath \,\ov\jmath \,\ov\jmath 
              \,\ov\jmath \,\big)^{aa}
  \qquad , \qquad 
  R_{21} \;=\; {1\0 N} \int \( \ov{\ov\jmath} \,\ov\jmath \,
               \cl D \,\ov\jmath  \)^{aa} 
              \quad , \quad \\ \label{4.11}
     & & \hspace*{-2.3cm}
  R_{31}^{(1)} \,=\, {1\0 N} \!\int\!\( \,\ov{\ov\jmath} \,\ov\jmath 
            \,\ov\jmath \,\cl D \,\ov\jmath \)^{aa} \, ,  \,
  R_{31}^{(2)} \,=\, {1\0 N} \!\int\!\( \,\ov\jmath \,\ov{\ov\jmath} \,\ov\jmath 
                  \,\cl D \,\ov\jmath \)^{aa}  \, , \, 
     R_{32} \,=\, {1\0 N}\!\int\!\(\,\ov{\ov\jmath} 
     \,\ov\jmath \,\cl D \,\ov\6 \,\cl D \,\ov\jmath \)^{aa} \; . \quad
\eea % 4.9 bis 4.11
The first index on $R$ refers to the $n$ of that $Q_n$, which under
application of $\cl T$ produces the $R$ in addition. The second index
just denotes the number of $\cl D$'s contained. $R_{n\,\cdots}$ has
the same mass dimension as $Q_n$, namely $m^{2n +2}$. In passing, 
$\6,\,\ov\6,\,j$, $\cl D\,$ and $\d_j\,$ have mass dimension $m^1\,$,
$\ep$ and $\int$ have $m^{-2}$ and $\cl T$ is dimensionless.

\subsection{ Technicalities in applying \boldmath$\cl T$ to invariants} 

If the $\cl D$'s are made explicit, an invariant is a linear
combination of products. To begin with $\cl T_1$ or $\cl T_2\,$,
let the $\d_{j_r^a}$ in there act on one factor of such a product,
for example on $\6 \ov{\ov \jmath}\,$:
$$
  \int \o_r^a \d_{j_r^a} {1\0 N}\!\int^{\prime\prime} \! \( 
  \bullet\bullet\bullet_{\rpp} \;\; \6^{\prime\prime} \, \ov{\ov \jmath}_{\rpp} \; 
  \circ\circ\circ_{\rpp} \)^{cc} \;\;    
  \krumm{6}{3}{1.3}{ {-i\0 N} \!\int \o_r^a \,f^{cba} \!\int^{\prime\prime}\!\! 
    \( \6^{\prime\prime} \ov\6^{\prime\prime \, 2}
   \, \d (\vc r - \vc r^{\prime\prime})  \)
   \( \circ\circ\circ \; \bullet\bullet\bullet \)_{\rpp}^{bc} }
   % Ende krumm-Inhalt       - - -  Nun oben rechts :
   \;\;\;\; {-i\0 N} \int \o_r^a \d_{j_r^a}\, f^{cbd} \!\int^{\prime\prime}\!\! 
   \( \6^{\prime\prime} \, \ov{\ov \jmath}_{\rpp}^{\,d} \)
   \( \circ\circ\circ \; \bullet\bullet\bullet \)_{\rpp}^{bc} $$ \\[2mm]
\bea{4.12}
   & &  \hspace*{4mm}
       \6^{\prime\prime} \to - \6 \;\,\hbox{(on $\d$)} \;\to +\6 \;
       \,\hbox{(by partial int.)\ , and similar with $\ov \6$ \ :} 
       \nonu \\[-12mm]
   \glu{18}
   {-i\0 N}\!\int \( \6 \ov\6^2  \o_r^a \) \,f^{cba} 
   \( \circ\circ\circ \; \bullet\bullet\bullet \)_r^{bc}    
   \,=\, {1\0 N}\!\int \( (\6 \ov\6^2 \o ) 
   \circ\circ\circ \; \bullet\bullet\bullet \)_r^{cc} \quad , \quad \hspace*{2.4mm}
\eea % 4.12
where $ -i f^{cba} \o^a \glr \o^{cb}\,$ was defined. To check \eq{4.12}\,:
if $\o$ were $j\,$, the operator $\int j^a\d_{j^a}$ would just count the 
number of $j$'s in the product, indeed. Remember that $\o^a$ is the sum
$\o^a = \o^{\heartsuit a} + \o^{\diamondsuit a}$ with $\o^{\heartsuit a}$
given by \eq{3.8}. But $\o^{\diamondsuit a}$ is rather taken from
\eq{4.18} below. Since $\o^a = \o^{\diamondsuit a} + O(\ep)\,$ and 
$L^{uv}_{00} = j^{uv} +O(\ep)\,$ we have $\o^a=j^a+O(\ep)\,$.

To study the action of $\,\cl T_3\,$ on an invariant, two factors in a 
product might be made explicit, for example $\ov \jmath$ and $\6 \ov \jmath\,$:
\be{4.13}
  \!\int\!\!\!\int^{\prime}\!\O_{r r'}^{a b}
  \,\d_{j_r^a}\,\d_{j_{\rp}^b}
  {1\0 N}\!\int^{\prime\prime}\!\! \( 
  \bullet\bullet\bullet_\rpp
  \;\; \ov\jmath_{\rpp} \; 
  \circ\circ\circ_{\rpp} \;\; \6^{\prime\prime} \,\ov\jmath_{\rpp} 
  \diamond\diamond\diamond_{\rpp} \)^{cc} \quad . \quad
\ee % 4.13
Here $\O_{r r'}^{a b}$ may be replaced by 
$\O_{r r'}^{a b} +\O_{r' r}^{b a} \glr \cl K_{r r'}^{ab}$, if the two
$\d$'s are applied ordered, i.e. interchanging them is forbidden. 
Assume $\ov\jmath_{\rpp}$ carries indices$^{uv}$ at its position
in $(\;\;)^{cc}\,$. Then
$\d_{j_r^a}\ov\jmath_{\rpp} =-if^{uva} \ov\6^{\prime\prime}
\d\(\vc r -\vc r^{\prime\prime}\) = i f^{uav} \cdots\,$.
We may relax the notation by $f^{uav} \glr f^a\,$ since the
indices $u,v$ are fixed by the position of $f^a$ in $( \;\; )^{cc}\,$.
Hence
\bea{4.14}
  \!\eq{4.13} \!\glo{7}\! {-1\0 N} \!\int\!\!\int^\prime\!\!\!\int^{\prime\prime}\!\!\!
     \cl K_{r r'}^{ab} \( \ov\6^{\prime\prime}
     \d (\vc r -\vc r^{\prime\prime} ) \)\,\(\6^{\prime\prime}
         \ov\6^{\prime\prime} \d (\vc r^\prime -\vc r^{\prime\prime}) \)
 \( \bullet\!\bullet\!\bullet f^a \!\circ\!\circ\!\circ f^b  
      \!\diamond\!\diamond\diamond \)_{\rpp}^{cc} \nonu \\
    \gluo{7}{7} {-1\0 N} \!\int\!\!\int^\prime\! \( \ov\6\,\6^\prime\,\ov\6^\prime
   \cl K_{r r'}^{ab} \) \,\d (\vc r -\vc r^{\prime})\,
     \( \bullet\bullet\bullet\, f^a \circ\circ\circ\, f^b  
     \diamond\diamond\diamond \)_r^{cc} \nonu \\
    \glu{7} {-1\0 N} \!\int\! \lb \ov\6\,\6^\prime\,\ov\6^\prime
        \cl K_{r r'}^{ab}\rb \( \bullet\bullet\bullet\, f^a 
        \circ\circ\circ\, f^b  
        \diamond\diamond\diamond \)_r^{cc} \quad . \quad 
\eea % 4.14
The pair of curly brackets stands for the coincidence limit  
$\lb any_{\,r\rp} \rb \gll \lim_{{\vcsm r}^\prime \to \vcsm r} 
any_{\,r\rp}\,$.

Imagine a definite invariant with its linear--combined products,
each booked down with all combinations of $f^a$ and $f^b$
(the latter to the right of the first). Let each of the various
terms be given the form \eq{4.14}. So far things are done by
hand on paper. But now it is convenient to put the result
(for $\cl T_3\,${\small\sl Invariant}) in a MAPLE--file and continue 
with it by keyboard and screen. There the following 5 steps
are done. 

{\bf I.}\, Reintroducing $\O$ by $\,\cl K_{r r'}^{ab} = \O_{r r'}^{a b} 
+\O_{r' r}^{ba}\,$ means that each $\O_{r r'}^{a b}$ becomes a partner with 
interchanged indices and variables (hence $\6$'s become primed and 
$\6^\prime$'s lose their primes). 

{\bf II.}\, With reason explained in step {\bf V}, convert all
   unprimed $\6$'s inside $\lb \,\, \rb$ into primed ones by using 
\be{4.15}
   \lb \6 \cdots \rb = \6 \lb \cdots \rb - \lb \6^\prime \cdots \rb
   \quad \hbox{and} \quad
   \lb \ov\6 \cdots \rb 
   = \ov\6 \lb \cdots \rb - \lb \ov\6^\prime \cdots \rb
   \quad . \quad
\ee % 4.15
To understand \eq{4.15} note that $\int \lb \6 \cdots \rb any_r
= \int\!\!\int^\prime\! (\6 \cdots ) \,\delta(\vc r - \vc r^\prime)\,
 any_r\,$. Now partial integrate under $\!\int$, use 
$\6\,\d = -\6^\prime \,\d$ and partial integrate with $\6^\prime$ 
under $\!\int^\prime$. Hence $\!\int \lb \6 \cdots \rb any_r$
$ = - \!\int \lb \6^\prime \cdots \rb any_r - \int 
\lb \cdots \rb \6\, any_r \,$, which leads to \eq{4.15} by a 
last partial integration.

{\bf III.}\, Partial integrate the unprimed $\6$'s in front of $\lb\;\;\rb\,$.
           Thereby some of the $( \quad )^{cc}$'s are differentiated,
           but they remain $( \quad )^{cc}$. Note the $1/N$--prefactor 
           of each invariant. Hence a typical term has the form 
\be{4.16}  \! 
 {1\0 N}\!\int\! \(..f^a..f^b..\)^{cc}
         \lb \hbox{\ft primed $\6$'s}\;\; \O \rb^{ab}
 = {1\0 N^2}\!\int\! \(..f^a..f^b..\)^{cc}
     \lb \hbox{\ft primed $\6$'s}\;\; (\6 -j ) \,\Theta \rb^{ab} 
     \,\; , 
\ee % 4.16
where \eq{2.10} was inserted.

{\bf IV.}\, Use \eq{4.15} for the last $\6$ in \eq{4.16} and partial
           integrate if it appears in front of $\lb \;\; \rb$.

{\bf V.}\, It will be shown in the sequel that $\Theta$ can be written as
\be{4.17} 
   \Theta_{r \rp}^{ab} \;=\; \sum_{s=0}^\infty \,\sum_{t=0}^\infty
   \,{\s^s\0 s!} \, {\ov \s^t \0 t!}\;\,L_{st}^{ab} \quad , \quad   
\ee % 4.17
with the coefficients $L$ depending on only the unprimed $z$
and $\ov z$. According to \eq{3.13} primed variables 
occur in $ \Theta_{r \rp}$ , \eq{3.14}, only through  
$\s =z^\prime -z$ and $ \ov \s =\ov z^\prime -\ov z\,$. This was 
the reason, to favour primed differentiations in
point {\bf II.}\, A coincidence limit makes $\s$ and $\ov \s$ vanish.
So, it depends on the $\6^\prime$'s and $\ov\6^\prime$'s in front
of $\Theta$ which coefficient $L$ survives. Replace $\lb 
\,{\rm primed}\, \6's \,\;\Theta \rb$
by this coefficient. End of the 5 MAPLE--steps. 

An example for the result is 
shown in Appendix C. Copy the result for $L$'s to a separate 
MAPLE--file. Here $\cl T_1\,${\small\sl Invariant} and 
$\cl T_2\,${\small\sl Invariant}\, might be included, and 
the ready list (see Appendix D) of $L$'s can be inserted there.

By \eq{4.17} the \glqq second $\o$\grqq\ becomes explicit.
\eq{2.10} simply turns into
\be{4.18}
 \o_r^{\diamondsuit a} \; = \; {i\0 N}\,f^{auv}\,\Theta_{rr}^{uv}
     \; =\; {i\0 N}\,f^{auv}\,L_{00}^{uv}
\ee % 4.18

\vskip -3mm
\subsection{ Derivation of the coefficients \,\boldmath$L$} 

%\nopagebreak
\eq{3.21} sugests to distinguish two parts\,: $\Theta =$ $^1\Theta$ $
+^2\Theta$ with $^1\Theta$ the first term of \eq{3.21} and $^2\Theta$ 
the remaining two terms in \eq{3.21} containing $\cl A\,$. 
Correspondingly $L$ splits into $^{1\!}L$ and $^{2\!}L$. To start with
\bea{4.19}
  ^1\Theta \glo{7} {1\0 \s } \( 1 - H_r \,\e^{ \s\,(-\ov \s/\ep 
          + \dup \,)}  H_{\zup \ov z}^{-1} \)
  = -{1\0 \s } H_r \sum_{s=1}^\infty {1\0 s!}\, \s^s (-\ov \s /\ep + \6 \,)^s
             H_r^{-1} \qquad \quad \nonu \\
  \gluo{6}{8} - \sum_{s=1}^\infty \sum_{t=0}^s {1\0 t!\,(s-t)!}\, 
           \s^{s-1} (-\ov \s /\ep)^t
           H_r \, \6^{s-t} H_r^{-1}
           \quad , \quad s\to s+1 \, : \qquad \nonu \\
  \glu{9}  \sum_{s=0}^\infty \sum_{t=0}^{s+1} \,{\s^s\0 s!}\,
        {\ov \s ^t \0 t!}\;\,
     { s! \0 (s +1 -t)! } { (-1)^{1+t} \0 \ep^t } B(s+1-t) \quad . \quad
\eea % 4.19
Defining $\theta_{t \leqslant s+1}$ to be 1 for $t \leqslant s+1$ and
zero otherwise $^{1\!}L$ is obtained as
\be{4.20}
  ^{1\!}L_{st} \; = \; \theta_{t \leqslant s+1}\; { s!\, (-1)^{1+t}\,
 B(s+1-t) \0 (s +1 -t)!\;\, \ep^t } \quad . \quad
\ee % 4.20

Towards $^{2\!}L$ let us assume that $\cl A$ can be written as
\be{4.21}
   \cl A^{ab} = \sum_{s=0}^\infty \,\sum_{t=0}^\infty
   \,{\s ^s\0 s!} \, {\ov \s ^t \0 t!}\;\, C_{st}^{ab} \quad , \quad   
\ee % 4.21
in general where the coefficients $C$ do not depend on $z^\prime$,
$\ov z^\prime\,$. If \eq{4.21} can be reached, $^{2\!}L$ can be 
traced back to $C$ as follows\,:     
\be{4.22}
   ^2\Theta = \6^\prime \cl A + \cl A \,j_\rp =
   \sum_{s=1}^\infty \,\sum_{t=0}^\infty
   \,{\s ^{s-1}\0 (s-1)!}\,{\ov \s ^t \0 t!}\;\, C_{st}^{ab}    
   + \cl A \sum_{u=0}^\infty {\s^u \0 u!} \sum_{v=0}^\infty 
        {\ov \s ^v \0 v!}
   \, \6^u \ov \6^v j_r \quad . \quad 
\ee % 4.22
Combining \eq{4.22} with \eq{4.21} one has to manipulate
\be{4.23}
  \sum_{s=0}^\infty \sum_{u=0}^\infty {\s^{s+u} \0 s!\,u!} \,=\,
  \sum_{u=0}^\infty \sum_{s=u}^\infty {\s^s \0 (s-u)!\,u!} \,=\,
  \sum_{s=0}^\infty {\s^s \0 s!} \sum_{u=0}^s {s!\0 u! (s-u)! } \quad . \quad
\ee % 4.23
Doing the same with the sums over $t$ and $v$ one can read off that
\be{4.24}
 ^{2\!}L_{st} = C_{s+1 \;t} + \sum_{u=0}^s \( \matrix{ s \cr u \cr} \)
   \sum_{v=0}^t \( \matrix{ t \cr v \cr} \) C_{s-u \,\; t-v} 
   \,\; \6^u \ov \6^v j_r \quad . \quad 
\ee % 4.24

It remains to determine $C_{st}\,$ as defined by \eq{4.21}.
Clearly ${\cl A}$ from \eq{3.24} needs a further rearrangement
such that $C_{st}\,$ can be read off. The two lines of \eq{3.24}
have something in common, namely the two sums over $p$ and $q$
followed by $1\0 p!\, q$ and explicit powers $p$ and $q\,$.
In the following formula $\a = - \s 
\6^\prime$\lower 2mm \hbox{$\!\!\!\!\!$\tiny$\bullet$\quad} ,  
$\b = -{\ep\0\s} \ov \6^\prime$\lower 2mm 
\hbox{$\!\!$\tiny$\uparrow$\quad}
apply to the first line of \eq{3.24} while 
$\a = - \s \6^\prime$\lower 2mm \hbox{$\!\!\!\!$\tiny$\bullet$\quad}
$+\s \ov \s /\ep$ and $\b =- {\ep \0 2\s} \ov\6^\prime$\lower 2mm 
\hbox{$\!\!\!$\tiny$\uparrow$\quad}$+{1\02}$ apply to the second\,:
\bea{4.25}
  \sum_{p=1}^\infty {1\0 p!}\,\a^p \sum_{q=1}^p {1\0 q}\,\b^q 
  &=& \int_0^1 \! d\tau  \sum_{p=1}^\infty {1\0 p!} \a^p \,\b\; 
       {1 - (\tau \b)^p \0 1-\tau\b \;} 
  = \int_0^1 \! d\tau \; {\b \( e^{(\tau\b -1)\a} -1 \) 
    \0 \tau\b -1}\,e^\a   \nonu \\ 
  &=& \; \sum_{c=1}^\infty {1\0 c! \; c} \,
   \Big( (\a\b-\a)^c - (-\a)^c \Big)\;e^\a \quad . \quad
\eea % 4.25
The last expression was obtained by expanding the exponential
(sum over $c$) and performing the $\tau$ integration.

The explicit exponential $e^\a$ at the right end of \eq{4.25} 
has welcome effects. It removes dot-subscripts by
$e^{-\s \dup^\prime} H_{\zup^\prime \ov z^\prime}  = 
H_{z \ov z^\prime}\,$ and
$e^{-\s \dup^\prime} H_{\zup^\prime \ov z}^{-1}
H_{\zup^\prime \ov z^\prime}
H_{z^\prime \ov z^\prime}^{-1}\,$  in the first and second line 
of \eq{3.24}, respectively. Moreover the second term
in the second-line-$\a$ compensates the prefactor
$e^{-\s \ov\s /\ep}$ in \eq{3.24}\,.

The corresponding intermediate result is
\bea{4.26}
  \cl A &=& \sum_{c=1}^\infty {1\0 c! \; c} \;\Bigg\{
  \( \big[ \s + \ep \ov \6^\prime \big]^c -\s^c \) 
           \6^c H_{z \ov z^\prime} H_\rp^{-1} \hspace*{7cm}\nonu \\
   & & \hspace*{18mm} - H_r {1\0 2^c} 
       \( \big[ \s + \ep \ov \6^\prime \big]^c -(2\s )^c \)
       \( \6 -{\ov \s \0 \ep} \)^c 
   H_r^{-1} H_{z \ov z^\prime} H_{z^\prime \ov z^\prime}^{-1}
  \Bigg\} \quad . \quad \\[-4.3mm] 
   & & \hbox{\tiny \hspace*{4.4cm} $\uparrow$ \hspace*{5.3cm}$\uparrow$
               \hspace*{6mm}$\uparrow$}     \nonu
\eea % 4.26   
where the arrow--subscript prevents $\ov \6^\prime$ to act on 
the explixit $\ov \s\,$. 

There is still some cumbersome analysis left. We first separate
powers of $\s$ and $\ov \s$ binomically\,: 
\bea{4.27}
  \cl A &=&
  \sum_{c=1}^\infty \sum_{k=1}^c {\ep^k \,\s^{c-k} \0 c\,k!\, (c-k)!} \;\cl B_1
  \;\;-\;\;
  \sum_{c=1}^\infty \sum_{k=1}^c \sum_{\ell=0}^c {\ep^{k-\ell}\, 
       \s^{c-k} c!\,(-\ov \s )^\ell \0 c\,k!\,(c-k)!\,
       \ell!\,(c-\ell)!\,2^c} \;\cl B_2 \qquad  \nonu \\  & &
  +\;\sum_{c=1}^\infty \sum_{\ell=0}^c { \s^c (-\ov\s )^\ell 
   (1 - 2^{-c}) \0 c \;\ell!\, (c -\ell)! \, \ep^\ell} 
   \;\cl B_3 \qquad \hbox{with} \quad
\eea % 4.27 
\be{4.28}
  \cl B_1 = \ov\6^{\prime \;k} \6^c H_{z \ov z^\prime} H_{\rp}^{-1} \;\; , \;\;
  \cl B_2 = H_r \ov\6^{\prime \; k} \6^{c-\ell}
   H_{z \ov z}^{-1} H_{z \ov z^\prime} H_{\rp}^{-1} \;\; , \;\;
   \cl B_3 = \cl B_2|_{k=0} \;\; . \quad 
\ee % 4.28
Now remember that primed indices $z^\prime$ and $\ov z^\prime$ 
must not occur in $C_{s\,t}\,$. With view to $z^\prime = z+\s\,$, 
$\ov z^\prime = \ov z + \ov \s$ they can be removed by Taylor 
expansions\,:
\bea{4.29}
 \cl B_1 \glo{7} \ov \6^k 
       \lower 2.2mm\hbox{$\!\!\!\!\!$\tiny$\uparrow\;\;$}
       \big(\6^c H_{z \,\ov z+\ov\s}   
       \lower 2.3mm\hbox{$\!\!\!\!\!\!\!\!$\tiny$\uparrow\;\quad$}
       \;\big) H_{z+\s \, \ov z+\ov\s}
       \lower 2.3mm\hbox{$\!\!\!\!\!\!\!\!$\tiny$\uparrow\;$}
       \quad =\;
        \sum_{s=0}^\infty \sum_{t=0}^\infty {\s^s \0 s!} 
        {\ov \s ^t \0 t!}\;
        \ov \6^{t+k}\big(\6^c H_{z \, \ov z}\big)\; \6^s
        H_{z \,\ov z}^{-1}   \nonu \\
  \glu{8} \sum_{s=0}^\infty \sum_{t=0}^\infty {\s^s\0 s!} 
       {\ov \s ^t \0 t!}
       \;M(0\,,\,t+k\,,\,c\,,\,s) \quad . \quad
\eea % 4.29
\vskip -5mm
Remember $M(a,b,c,d)$ from \eq{3.9}. Furthermore
\bea{4.30}
 \cl B_2 \glo{9} \sum_{s=0}^\infty \sum_{t=0}^\infty 
            H_r \,\dup^{c-\ell} H_{\zup \ov z}^{-1}\, {\ov \s ^t \0 t!} 
            \ov \6^{t+k} H_{\zup \ov z}\, {\s^s \0 s!} \,
            \6^s H_{z \ov z}^{-1} \;\; , \;\; \nonu \\
  \gluo{6}{8} \sum_{s=0}^\infty \sum_{t=0}^\infty {\s^s\0 s!} 
        {\ov \s ^t \0 t!}
      \sum_{a=0}^{c-\ell} {(c-\ell)! \0 a! \; (c-\ell-a)!}
      H \( \6^{c-\ell-a} H^{-1} \) \ov\6^{t+k}\( \6^a H \) \6^s H^{-1}
      \nonu \\
  \glu{9} \sum_{s=0}^\infty \sum_{t=0}^\infty {\s^s\0 s!} 
       {\ov \s ^t \0 t!}
      \,\sum_{a=0}^{c-\ell} {(c-\ell)! \0 a! \; (c-\ell-a)!}
      \;M(c-\ell-a\,,\,t+k\,,\,a\,,\,s) \quad . \quad 
\eea % 4.30
Inserting \eq{4.29},\eq{4.30} and $\cl B_3 = \cl B_2|_{k=0}$  
into \eq{4.27} we are forced to commute a few double sums  
to reach the structure \eq{4.21}\,. The final result is 
\bea{4.31}
  C_{s\,t} 
  &=& \sum_{c=0}^s {s! \0 (s-c)!} \sum_{k=1}^\infty \;\Bigg(\;
    {\;\ep^k \; M(0,t+k,c+k,s-c) \0 c!\,k!\,(c+k) } 
    \hspace{4.7cm} \nonu \\[1.5mm]
  & & \hspace*{-1cm}
  -\!\!\!\!\sum_{\ell=0}^{{\rm min} (t,c+k)}
   \!\!\!\!\!{t! \0 (t-\ell)!} \sum_{a=0}^{c+k-\ell}
       {\;\ep^{k-\ell} \; (-1)^\ell (c+k-1)!\, 
          M(c+k-\ell-a,t+k-\ell,a,s-c) \0 
       k!\;c!\;\ell!\;a!\;(c+k-\ell-a)!\; 2^{c+k} } 
        \,\;\Bigg)  \nonu \\[1.5mm]
  & & \hspace*{-1.8cm}
    + \,\theta_{s\geqslant 1} \sum_{c=1}^s {s! \0 (s-c)!} 
        \!\!\sum_{\ell=0}^{{\rm min} (t,c)}
      \sum_{a=0}^{c-\ell} {t! \0 (t-\ell)!} \;{(1-2^{-c})
          \,(-1)^\ell \,M(c-\ell-a,t-\ell,a,s-c) \0
         \ep^\ell \;\;c \;\ell!\; a!\; (c-\ell-a)!} \quad . \;\;
\eea % 4.31

We shall work with \eq{4.31} neither by pen nor on sreen. But it
is an easy task to write \eq{4.31} into a file. MAPLE performs
the sums for index pairs ${s\,t}$ of interest. It also processes
$^{2\!}L_{s\,t}\,$, \eq{4.24}\,, and adds $^{1\!}L_{s\,t}\,$,
\eq{4.18}\,. The coefficients $M(a,b,c,d)$ appearing in $L$ can be 
made explicit by hand on screen and read in. Hence all $L_{s\,t}$
of interest are known and expressed by the currents $j$. 
A list of some $L$'s is shown in Appendix D.

We are quite sure that every detail of this section 4.3 is correct.
This rests on various successful tests. $\cl T$ applied to a
holomorphic invariant must result in a linear combination of
holomorphic invariants. Any erraneous little detail in e.g.\ \eq{4.31}
(and that had happened) would lead to the desaster of
non--holomorphic remnants in the result (and that had led to
find the error). 

The above section is the last one which avoids approximations or
truncations. The kinetic energy operator $\cl T$ \anke see \eq{2.8}, 
\eq{3.8},\eq{4.18}, \eq{4.17} and all about $C$'s and $L$'s \anke 
contains the $\ep$--powers to all orders. It remained general. 
The sad counterpart is that the whole apparatus developed above 
will by far not be exhausted in what follows. \\[3mm]
    \hspace*{15mm} \rule[1mm]{19mm}{.1mm} \, {\ft\sl 
    end of the general analysis that keeps $\ep$ arbitrary} 
    \, \rule[1mm]{19mm}{.1mm} \\[-8mm]

\let\dq=\thq \renewcommand{\theequation}{5.\dq}
                           \setcounter{equation}{0}
\section{ Results of {\boldmath$\cl T$}--application}
\vskip -4mm
\bea{5.1}
  \cl T \,Q_0 &=& -{\aleph \0 2\,\ep^2} +{13\0 8}\,Q_0 + \ep\,{5\0 4}\,Q_1 
    + \ep^2\({19\0 384}\,Q_2 + {1\0 48}\,R_{21} -{5\0 32}\,R_{20} \) \nonu \\
   & & +\,\ep^3 \({7\0 576}\,Q_3 + {35\0 288}\,R_{32} -{77\0 144}\,R_{31}^{(1)}
       -{49\0 144 }\,R_{31}^{(2)} \) \; + \cl O\(\,\ep^4\,\) \quad . \quad 
   \\[2mm] \label{5.2}
  \cl T \,Q_1 &=& +{\aleph \0 2\,\ep^3} +{69\0 32}\,Q_1 +\ep \(
          {149\0 128}\,Q_2 -{11\0 8}\,R_{21} +{3\0 32}\,R_{20} \) \nonu \\
       & & +\,\ep^2 \( {33\0 512}\,Q_3 -{27\0 256}\,R_{32} 
           +{165\0 128}\,R_{31}^{(1)} +{105\0 128}\,R_{31}^{(2)} \)
           \; + \cl O\(\,\ep^3\,\)   \quad . \quad
   \\[2mm] \label{5.3}
  \cl T \,Q_2 &=& -{3\,\aleph \04\,\ep^4} +{1\0 8\,\ep^2}\,Q_0 
                  +{3\0 2\,\ep}\,Q_1 +{1943\0 768}\,Q_2 -{25\0 96}\,R_{21} 
                  +{7\0 64}\,R_{20} \nonu \\
        & & +\,\ep\( {35\0 32}\,Q_3 -{43\0 16}\,R_{32}    
              -{13\0 8}\,R_{31}^{(1)} -{11\0 8}\,R_{31}^{(2)} \)
              \; + \cl O\(\,\ep^2\,\)     \quad . \quad
   \\[2mm]    \label{5.4}
  \cl T \,Q_3 &=& +{3\,\aleph\0 2\,\ep^5} +{3\0 8\,\ep^3}\,Q_0 -{103\0 32\,\ep^2}\,Q_1
             +{1\0 \ep}\,\( {259\0 64}\,Q_2 +{43\0 32}\,R_{21} \) \nonu \\
      & & +\,{2749\0 1024}\,Q_3 -{1333\0 1536}\,R_{32} +{1403\0 768}\,R_{31}^{(1)}
           +{1111\0 768}\,R_{31}^{(2)}  \; + \cl O\(\,\ep\,\) \quad . \quad
   \\[3mm] \label{5.5}
   \cl T \,R_{20} &=& -{5\0 2\,\ep^2}\,Q_0 +{13\0 4}\,R_{20} 
        +\,\ep\( -{25 \0 8} R_{31}^{(1)} -{7 \0 4} R_{31}^{(2)} + {9\0 4}\, S \) 
         \; + \cl O\(\,\ep^2\,\) \;\; . \quad
   \\[2mm] \label{5.6}        % H37 , J48
   \cl T \,R_{21} &=& +{1\0 8\,\ep^2}\,Q_0 +{3\0 16\,\ep}\,Q_1 -{3\0 64}\,Q_2
                +{45\0 16}\,R_{21} -{13\0 32}\,R_{20} \nonu \\
    & &  +\,\ep\( {29\0 32}\,R_{31}^{(1)} -{13\0 16}\,\,R_{31}^{(2)} \)
            \; + \cl O\(\,\ep^2\,\)  \quad . \quad
  \\[2mm] \label{5.7}     % J46
   \cl T \,R_{31}^{(1)} &=& -{1\0 2\,\ep^3}\,Q_0 +{1\0 2\,\ep^2}\,Q_1
          -{1\0 8\,\ep}\,R_{21} \nonu \\
  & & +\,{1\0 32}\,R_{32} +{133\0 32}\,R_{31}^{(1)}
       +{17\0 64}\,\,R_{31}^{(2)} -{3\0 4}\,S  \; + \cl O\(\,\ep\,\)
        \quad . \quad
   \\[2mm] \label{5.8}       % J46
   \cl T \,R_{31}^{(2)} &=& -{1\0 4\,\ep^3}\,Q_0 +{1\0 4\,\ep^2}\,Q_1
          -{1\0 4\,\ep}\,R_{21} \nonu \\
    & & -\,{1\0 16}\,R_{32} -{1\0 8}\,R_{31}^{(1)}
        +{7\0 2}\,\,R_{31}^{(2)} +{1\0 4}\,S \; + \cl O\(\,\ep\,\)
        \quad . \quad
  \\[2mm] \label{5.9}
   \cl T \,R_{32} &=& +{1\0 4\,\ep^2}\,Q_1 +{1\0 8\,\ep}\,Q_2 
                      +{5\0 8\,\ep}\,R_{21} \nonu \\
   & & -\,{7\0 384}\,Q_3 +{637\0 192}\,R_{32} +{125\0 96}\,R_{31}^{(1)}
             +{35\0 48}\,R_{31}^{(2)}  \; + \cl O\(\,\ep\,\) \quad . \quad
\eea % 5.1 bis 5.9

Obviously, on the left hand sides $\cl T$ is applied to all the
invariants named in \eq{4.9} to \eq{4.11}. The necessity to include
$\ep$--powers diminishes with increasing index on $Q$ or left on $R\,$.
On the right hand sides there are two unknown quantities, which
have to be invariants too. The most simple holomorphic invariant
of the form \eq{4.8} is obtained by choosing \glqq $any$\grqq$=1\,$:
\be{5.10}
\aleph = {1\0 N} \int \big( \; 1 \; \big)^{aa} = {N^2-1\0 N} \,F 
         \qquad\quad 
\ee % 5.10
with $F$ the divergent area of the plane. The equation 
$\cl T \,\cl P  = \cl E - \cl V$ to be solved shows, that $\aleph$--terms
can be absorbed in $\cl E\,$. The quantity $S\,$
\be{5.11}
   S \gll {1\0 N^2} \int \(\; f^a \;\ov\jmath\;\ov\jmath\;
   f^a \,\ov{\ov \jmath}\, \cl D \ov\jmath \; \)^{cc}  \quad , \quad
\ee % 5.11
is also a holomorphic invariant (sum over $a$). To check this
invariance, use $h^{ab} = 2\Sp \( T^a h T^b h^{-1}\)\,$,
$f^{eas} T^s = -{\rm i}\lk T^e,T^a\rk$ and $2\Sp( \cdots T^e )
\,2\Sp(T^e \cdots) = 2\Sp(\cdots)$ to get $(h^{-1} f^a h)^{cr} =
f^{cdr} h^{ad}\,$. Hence the integrand $\big( \quad \big)^{cc} \glr 
S_{\rm int}$ of $S$ transforms into 
\bea{5.12} \! 
 S_{\rm int} &\!\!\to\!\!& ( h^{-1} f^a h )^{cr}\; (\ov\jmath\;\ov\jmath )^{rs}\;
 ( h^{-1} f^a h )^{st}\; (\,\ov{\ov \jmath}\,\cl D \ov\jmath \; )^{tc} \qquad \nonu \\
 &\!\!=\!\!& f^{cdr} (h^{-1})^{da} \; (\ov\jmath\;\ov\jmath )^{rs}\; 
       f^{sgt} h^{ag} (\,\ov{\ov \jmath}\,\cl D \ov\jmath \; )^{tc}
  \,=\,  \( \; f^g \;\ov\jmath\;\ov\jmath\;
    f^g \; \ov{\ov \jmath}\,\cl D \ov\jmath \; \)^{cc} = S_{\rm int}
  \;,\,\; \hbox{q.e.d.} \;\;\quad 
\eea % 5.12

All invariants occurring on the right hand sides of \eq{5.1} to \eq{5.9}
migth also appear under $\cl T$--operation on the left. The next 
section will explain why. $\cl T \aleph = 0$ is no problem. However 
through the evaluation of $\cl T S\,$ some unsurmountable difficulties
will become obvious in Section 6.3.

\let\dq=\thq \renewcommand{\theequation}{6.\dq}
                           \setcounter{equation}{0}
\section{ The first steps in solving \boldmath$\,\,\cl T \,\;\cl P \; 
           = \;\cl E - \cl V\,$}

The solution $\cl P$ to this equation is some function of $\ep\,$.
Only after this function is obtained the limit $\ep \to +0$
can be performed. This requirement is violated from the outset
if we truncate the {\sl ansatz} for $\cl P$ at some highest power
$n$ of $\ep$ (\glqq step $n$\grqq ). Let us nevertheless do so
\anke with thoughts on truncated perturbative expansions.     

In the most simple $\cl P$--{\sl ansatz} let no positive $\ep$-power
be allowed at all\,: {\bf step zero}\,. Remember that $\cl V$ equals 
$Q_0\,$ (combine \eq{1.7} with \eq{1.1}\,). With the
\be{6.1}
   \hbox{{\sl ansatz }} \qquad  \ \cl P = c_0\; Q_0 \qquad . \quad
\ee % 6.1
The equation reads $c_0 \cl T Q_0 = \cl E - Q_0\,$. Omitting the
positive $\ep$ powers in \eq{5.1} one obtains
\be{6.2}
     c_0 = -{8\0 13} \qquad  \hbox{and} \quad 
            \cl E = {4\0 13\,\ep^2}\,\aleph \qquad . \quad
\ee % 6.2
The result $\cl P = -{8\0 13}\,Q_0$ differs from \eq{1.8} i.e. from
$-{1\0 2}\,Q_0$ \,, \anke the well known misfortune.

\vskip -3mm
\subsection{ Step one\ : \ including \boldmath$\ep^1$} 

Now $\cl P$ must contain $\ep^0$ and $\ep^1$. The mass dimension $m^2$
of all terms in $\cl T \,\cl P = \cl E - Q_0$ is dictated by the
inhomogeneity $Q_0$ ($\ep$ has $m^{-2}$, $Q_1$ has $m^4\,$). So we are
lead to the
\be{6.3}
   \hbox{{\sl ansatz }} \qquad  \ \cl P = c_0\;Q_0 + c_1\;\ep\,Q_1 
   \quad  \quad
\ee % 6.3
and expect two equations for the coefficients by comparison.
With \eq{5.1} and \eq{5.2} one obtains
\be{6.4}
  \(-c_0 +c_1\)\,{\aleph \0 2\,\ep^2} +c_0\,{13\0 8}\,Q_0 +
   \ep\,\(\,c_0\,{5\0 4} + c_1\,{69\0 32} \)\,Q_1 
   = \cl E - Q_0 \quad , \quad 
\ee % 6.4
and hence
\be{6.5}
   c_0 = -{8\0 13} \quad , \quad  c_1 = {320 \0 13*69} 
       \qquad \hbox{and} \quad
       \cl E = {\aleph \0 2\,\ep^2}\,\({320\069*13} + {8\0 13}\) 
       \quad . \quad    
\ee % 6.5
The limit $\ep \to +0$ \glqq afterwards\grqq\ reduces $\cl P$ to
$ -{8\0 13}\,Q_0$ again. No change.

\vskip -3mm
\subsection{ Step two\ : \ including \boldmath$\ep^2$} 

$\cl P$ must contain $\ep^0\,$, $\ep^1\,$ and $\ep^2\,$. 
Already \eq{5.1} shows that now the invariants $R_{21}$ and $R_{20}$
come into play. It will be seen that their inclusion has influence
on the $Q_0$--prefactor in question. Inspecting \eq{5.5} and \eq{5.6}
one may expect a closed set of equation. However five coefficients
are to be distinguished\,: 
\be{6.6}
   \hbox{{\sl ansatz }} \qquad  \ \cl P = c_0\;Q_0 \,+ c_1\;\ep\,Q_1 
  \,+ c_2\;\ep^2\,Q_2  \,+ c_3\;\ep^2\,R_{21} \,+ c_4\;\ep^2\,R_{20}   
   \quad . \quad
\ee % 6.6
Now the $\cl T$--applications \eq{5.1} to \eq{5.3} and \eq{5.5},
\eq{5.6} are in use. We compare the coefficients of $Q_0\,$,
$\ep Q_1\,$, $\ep^2 Q_2\,$, $\ep^2 R_{21}\,$ and $\ep^2 R_{20}\,$.
The resulting 5 equations can be read off from the following
MAPLE program\,:
\bea{6.7}
  {\rm resu} \!&:=&\! {\rm solve}
    \big(\;\big\{\; 13*c_0 + c_2 + c_3 -20*c_4 = -8\,, \hspace{4cm} \nonu \\
  & & \hspace{14mm} 40*c_0 + 69*c_1 + 48*c_2 + 6*c_3 = 0\,, \nonu \\
  & & \hspace{14mm} 38*c_0 + 894*c_1 + 1943*c_2 - 36*c_4 = 0\,, \nonu \\
  & & \hspace{14mm} 25*c_0 - 132*c_1 - 25*c_2 + 270*c_3 = 0\,, \nonu \\
  & & \hspace{-18mm} -10*c_0 + 6*c_1 + 7*c_2 - 26*c_3 + 208*c_4 = 0
  \;\big\}\,,\,\big\{\, c_0,c_1,c_2,c_3,c_4 \,\big\}\, \big)\;;  \quad . \quad
\eea % 6.7
Again the limit $\ep \to +0$ \glqq afterwards\grqq\ reduces $\cl P$ to
the first term, i.e. only $c_0$ remains of interest for $\cl P$\,:
\bea{6.8}
  \hbox{request} &:& c_0 = -0.5000 \hspace{5cm} \nonu \\
   \hbox{step 1} &:& c_0 = -0.6154 \,\; =\, -8/13 \nonu \\ 
   \hbox{step 2} &:& c_0 = -0.6330  \\
  \big(\; \hbox{no R's} &:& c_0 = -0.5986 \;\big) \nonu  
\eea % 6.8
Aside, the energy is $\,\cl E = ( -2\,c_0 +2\,c_1 -3\,c_3)
\;\aleph/(4\,\ep^2)\,$. \,Obviously, step 2 has led further away
from the \glqq request\grqq . But one should not expect too much from the
present lowest non--trivial order. The important outcome is merely {\bf
that} $c_0$ {\bf does change} under the inclusion of higher $\ep$ powers. 

In the last line of \eq{6.8} we were curious what happens when
$R_{21}$ and $R_{20}$ were deleted from the {\sl ansatz} and
from \eq{5.1} to \eq{5.3}. Then \eq{6.7} would reduce to the first
three lines with $c_3 = c_4=0\,$. So, the $R$'s {\sl \,d\,o\,} influence
the value of $c_0\,$. Stimulated by the above we might study one more
step at least.  

\subsection{ The problems with step three\ : \ including \boldmath$\ep^3$} 

Now $\cl P$ must include $\ep^0\,$ to $\ep^3\,$, and (at least)
all $\cl T$--applications \eq{5.1} to \eq{5.9} become relevant.
In addition $\cl T\,S\,$, see \eq{5.11}), has to be evaluated,
but this makes trouble\,:.  
\be{6.9}
 \cl T\, S = {1\0 2\,\ep^3}\,Q_0 -{1\0 2\,\ep^2}\, Q_1 
  +{1\0 16\,\ep}\,R_{21} -{1\0 2\,\ep}\,S^{(2)} \; + \cl O\(\,\e^0\,\)
 \quad . \quad
\ee % 6.9
In \eq{6.9} one more object appears. $S^{(2)}$ is a holomorphic invariant
too and reads
\be{6.10}
     S^{(2)} \gll {1\0 N^3} \int \( \; f^a \; \ov{\ov\jmath} \; f^b\;f^a\; 
     \ov\jmath \; f^b\; \cl D \ov\jmath \;\)^{cc} \qquad
     \hbox{(sum over $a$ and $b$)}\footnote{\,$S$ and $S^{(2)}$
     can be depicted as
\unitlength .8cm
  \hspace*{6mm}
  \begin{picture}(0,1)
    \put(1.5,0){\oval(2,1)} 
    \put(2.5,.26){\line(3,1){.5}}
    \put(2.5,-.26){\line(3,-1){.5}}
              \put(1.5,-.5){\line(0,1){1}}
      \put(.5,.26){\line(-3,1){.5}}
      \put(.5,-.26){\line(-3,-1){.5}}
  \put(3.2,-.55){$\cl D \ov\jmath$}
  \put(3.2,.3){$\ov{\ov\jmath}$}
    \put(-0.4,-.55){$\ov\jmath$}
    \put(-0.4,.3){$\ov\jmath$}
 \put(4.1,0){\hbox{and}}  %----------------------------
    \put(6.3,0){\oval(1,1)} 
        \qbezier(6.1,-.5)(6.33,-.02)(6.56,.46) % lang
        \qbezier(6.04,.46)(6.16,.22)(6.28,-.02)   % oben kurz
        \qbezier(6.32,-.12)(6.41,-.31)(6.5,-.5)  % unten kurz
     \put(6.3,.5){\line(0,1){.3}}
  \put(6.8,-.1){\line(1,0){.3}}
  \put(5.8,-.1){\line(-1,0){.3}}
      \put(6.1,.9){$\cl D \ov\jmath$}
      \put(5.2,-.2){$\ov\jmath$}
      \put(7.2,-.2){$\ov{\ov\jmath}$}
  \end{picture} \rule[-6mm]{0pt}{12mm}
  \hspace*{5.7cm} \, , respectively\,. } \quad . \quad
\ee % 6.10
It has mass dimension $m^6\,$ (see end of \S~4.1). $S$ and all terms 
in \eq{6.9} have mass dimension $m^8\,$. In \eq{6.9} the term 
$\cl O \(\,\ep^0\,\)$ is not made explicit (but should be). It is again 
an invariant and has the structure of $S^{(2)}$ with one more $\cl D$ in. 
Clearly, now $\cl T$ has to be applied to both of these new invariants.
It may be supposed that thereby even more holomorphic invariants
come into play. And so on, and so on \anke a desaster. As a warning, 
further steps in the desert become quite time consuming.

Is step three at its very end\,? There is a last chance.
We abandon $S$ from the {\sl ansatz}\,:
\be{6.11}
  \cl P = c_0 Q_0 +c_1 \ep^2 Q_1 +c_2 \ep^3 Q_3 
          +c_4 \ep^2 R_{21} +c_5 \ep^2 R_{20} +c_6 \ep^3 R_{32}
          +c_7 \ep^3 R_{31}^{(1)} +c_8\ep^3 R_{31}^{(2)} \;\; . \; 
\ee % 6.11
Then via \eq{5.5}, \eq{5.8}, \eq{5.9} a term 
$\,\Delta \gll (9c_5 -3c_7 +c_8)\ep^3 S/4\,$ remains in $\cl T \cl P\,$.
May we hope for the miracle that this term vanishes\,? \,MAPLE solves
the 9 equations for $c_0$ to $c_8$ with ease. It produces
$\,c_0 = -0.5562\,$ \ \anke \ and $\,\Delta =-0.275\,\ep^3 S\,
  \neq 0$\,: no miracle.

\let\dq=\thq \renewcommand{\theequation}{7.\dq}
                           \setcounter{equation}{0}
\section{ Conclusions}

 The kinetic energy operator $\cl T$ is worked out  as a power
 series in the regularizaion parameter $\ep\,$. $\cl T$ contains
 all powers of $\ep$ and their  prefactors are made explicit.
 $\cl T$ is applied to various holomorphic invariants and leads to
 linear combination of such invariants. Admittedly the detailed results
 include only the third power of $\ep\,$ at highest. 

 To obtain the ground state $\psi =e^P$ (with $P$ restricted to its 
 leading nontrivial term towards $m \to \infty$) the Hamilton-Jacobi 
 type equation $\,\cl T \,\cl P = \cl E -\cl V\,$ needs to be solved 
 for $\cl P$.  But actually only $\ep^1$ and $\ep^2$ could have been 
 included in an {\sl ansatz} for $\cl P$.  

 So far tested, the point splitting regularization works
 well and is thus hopefully freed from possible doubts. 
 Due to the difficulties with breaking apart the $\ep$ power
 series, one migth think about to sum it up. This could be
 possible, but then\,: who has the strength to guess a suitable 
{\sl ansatz} for $\cl P\,$?

%%%%%%%%%%%%%%%%%%%%%%%%%%%%%%%%%%%%%%%%%%%%%%%%%%
%%%%%%%%%%%%%%%%%%%%%%%%%%%%%%%%%%%%%%%%%%%%%%%%%%
%                Die ANHÄNGE
%%%%%%%%%%%%%%%%%%%%%%%%%%%%%%%%%%%%%%%%%%%%%%%%%%
%%%%%%%%%%%%%%%%%%%%%%%%%%%%%%%%%%%%%%%%%%%%%%%%%%
\newpage

\let\dq=\thq \renewcommand{\theequation}{A.\dq}
                           \setcounter{equation}{0}

\parag{Appendix A : \,\hbox{\small How \eq{2.5} and 
        \eq{2.8}\,, \eq{2.9}\,, \eq{2.10} come about}}

\rule[.5mm]{3cm}{.9mm}\, \eq{2.5} :  \\[1mm]
When restricting $\cl T$ to act in the space $\psi\!\lk j_r^a\rk$ and since
$p_{\rpp}^v$ is the operator at the right end of \eq{2.1} the first detail to be
studied is $p_{\rpp}^v \psi\!\lk j_r^a\rk\,$. With \eq{2.3} this reads
\be{A1}
   p_{\rpp}^v \psi =
   \6^{\prime\prime}( M_{\rpp}^{\!-1} )^{vc} 
  \d_{A^c(\vcsm r^{\prime\prime})} \, \psi  
  =  \6^{\prime\prime}\, M_{\rpp}^{cv} \int^{\prime\prime\prime}\!\!
      \( \d_{A^c(\vcsm r^{\prime\prime})} \,j_{r^{\prime\prime\prime}}^e \)
       \,\d_{j_{r^{\prime\prime\prime}}^e}  \psi  \quad . \quad 
\ee % A1
For the above round bracket one may combine $j^e = 2\Sp \( T^e (\6 H) H^{-1} \)$
with $H = M^\dagger M$ to get
$j^e = 2\Sp \big( T^e (\6 M^\dagger) M^{\dagger -1}
+ T^e M^\dagger (\6 M) M^{-1} M^{\dagger -1} \big)$. 
Now $(\6 M) M^{-1} = i T^b A^b$ (see e.g. p.22 of \cite{ich}) and the 
$A$-independence of $M^\dagger$ lead to 
\be{A2}
    \d_{A^c(\vcsm r^{\prime\prime})} \,j_{r^{\prime\prime\prime}}^e 
  = i (M_{\rpp}^\dagger)^{ec} 
    \d (\vc r^{\prime\prime} - \vc r^{\prime\prime\prime})
\ee % A2
Using \eq{A2} in \eq{A1} one arrives at \eq{2.5} in the main text.  	  

\rule[.5mm]{3cm}{.9mm}\, \eq{2.8} : \\[1mm]
As explained below \eq{2.7} the kinetic energy operator $\cl T$ splits
into three parts.  The first is $\cl T_1 = \int \o_r^{\heartsuit a}\,
\d_{j_r^a}$ and requires evaluation of
\bea{A3}
  \ov{p}_{\rp}^u \; {\cl G }^{bv}_{r \rpp} \glo{5}
   -\ov \6'\,( M_{\rp}^\dagger )^{ud} \d_{A^{d *}(\vcsm r')}  
   G_{r \rpp} \( \d^{bv} - e_{r \rpp} \( {H_{z'' \ov z}}^{\! -1} 
   \, H_{z'' \ov z''} \)^{bv} \; \) \quad , \quad \hspace*{2cm} \\[-4mm] 
   & & \ \ \hbox{\small where} \ e_{r \rpp}  \ \hbox{\small is shorthand for} \ \ 
       \exp\( -(\vc r -\vc r^{\prime\prime})^2/\ep\) \ , \quad \nonu \\[-4mm]   
             \label{A4}
   \glu{9} \ov \6'\,( M_{\rp}^\dagger )^{ud} G_{r \rpp} e_{r \rpp}
        \lk \( \cl R \d_{A^{d *}(\vcsm r')} H_{\rpp}^{cb} \) H_{\rpp}^{cv}
          + \( \cl R H_{\rpp}^{cb} \) \d_{A^{d *}(\vcsm r')} H_{\rpp}^{cv} \rk
          \quad , \qquad \\
   & & \ \ \hbox{\small where the operator $\cl R$ replaces 
           ${\ov z}^{\prime\prime}$ by $\ov z$ within its round bracket .}
       \quad  \nonu 
\eea % A3 und A4
To proceed with \eq{A4} we need
\be{A5}
     \d_{A^{d *}(\vcsm r')} H_{\rpp}^{cb} =
     {\ov G}_{\rpp \rp} (M_{\rp}^\dagger)^{gd}\,f^{cgh}\, H_{\rpp}^{hb} 
     \quad \quad
\ee % A5
and the same with $v$ in place of $b\,$. To obtain \eq{A5}
derive from $M^{\dagger -1} \ov{\6} M^\dagger = -i T^a A^{a*}$
a differential equation for $\d_{A^{d *}(\vcsm r')} M_{\rpp}^\dagger$
and solve it to get
\be{A6}
   \d_{A^{d *}(\vcsm r')} M_{\rpp}^\dagger
    = -i M_{\rp}^\dagger T^d M_{\rp}^{\dagger -1} \ov G_{\rpp \rp}
    M_{\rpp}^\dagger \ \  \hbox{or} \ \
   \d_{A^{d *}(\vcsm r')} H_{\rpp} 
    = -i \ov G_{\rpp \rp} M_{\rp}^\dagger T^d M_{\rp}^{\dagger -1}
    H_{\rpp} \ \  , \quad
\ee % A6
since in $H=M^\dagger M$ only $M^\dagger$ depends on $A^*\,$.
Now using this and $H_{\rpp}^{cb} = 2\Sp( T^c H_\rpp T^b H_{\rpp}^{-1}  )$
on the lhs of \eq{A5} one obtains
\be{A7}
   \d_{A^{d *}(\vcsm r')} H_{\rpp}^{cb} = -i \ov G_{\rpp \rp}
   2\Sp (T^c D B -T^c B D ) =
   -i \ov G_{\rpp \rp} 2\Sp (\lk T^h , T^c \rk\!\! D)\,2\Sp( T^h B)
   \ \ 
\ee % A7
with the temporary abbreviations $D = M_{\rp}^\dagger T^d M_{\rp}^{\dagger
-1}$ and $B = H_\rpp T^b H_{\rpp}^{-1}$. From \eq{A7} one derives the
rhs of \eq{A5} indeed.  

Using \eq{A5} in \eq{A4} we arrive at
\be{A8}
 {\ov p}_{\rp}^u \; {\cl G }^{bv}_{r \rpp} = 
  G_{r \rpp}\, e_{r \rpp} \, f^{uhc}\,H_{\rpp}^{hv} \,
  H_{z^{\prime\prime}\ov z}^{cb} \( {\ov \6}^\prime \,{\ov G}_{\rpp \rp} 
- \cl R \,{\ov \6}^\prime \, {\ov G}_{\rpp \rp} \)\, \quad . \quad
\ee % A8
Note that ${\ov \6}^\prime$ and $\cl R$ do commute. The round
bracket in \eq{A8} restricts their action. $\ov G$ is $\e$--regulated,
see \eq{2.4}. Due $\e \to 0$ the first term may be replaced by 
$ - \d\(\vc r^{\prime\prime} -\vc r^\prime\)\,$ but the second 
requires more care\,:
\bea{A9}
  - \cl R \,{\ov \6}^\prime \,\ov G_{\rpp \rp} \,=\,  
   {\e^2/\pi \0 \lk (z^{\prime\prime} -z^\prime)\,(\ov z - \ov z^\prime)
    + \e^2 \rki^2 } \,=\, 
   e^{( {\ov z}^{\prime\prime} -\ov z ) {\ov \6}^\prime} 
  \,{\e^2/\pi \0 
   \lk (z^{\prime\prime} -z^\prime)\,(\ov z^{\prime\prime} - \ov z^\prime)
    + \e^2 \rki^2 } \;\; , \;
\eea % A9
where the exponential is a translation operator ($\ov z^\prime \to
\ov z^\prime +\ov z^{\prime\prime} -\ov z\,$)\,.

The couple \eq{A8} (including \eq{A9}) is now well prepared for insertion
into \eq{2.6} at the right inner position. Thereby slightly rearranged 
the result $\cl T_1$ reads\,
\bea{A10}
 \cl T_1 &=& {i\pi\0 N} \!\int\!\!\int^\prime \!\int^{\prime\prime}
  G_{r \rpp}\, e_{r \rpp} \, f^{uhc} H_r^{ab} H_{z^{\prime\prime}
  \ov z}^{cb}\;\hbox{\Large \boldmath$($}\, -\ov{\cl G }^{au}_{r \rp}\,
  \d\(\vc r^{\prime\prime} -\vc r^\prime\) \nonu \\
  &+& \,{\e^2/\pi \0 
   \lk (z^{\prime\prime} -z^\prime)\,(\ov z^{\prime\prime} - \ov z^\prime)
    + \e^2 \rki^2 } \;
    e^{( \ov z -{\ov z}^{\prime\prime}) {\ov \6}^\prime}\,
    \ov{\cl G }^{au}_{r \rp}\;\,\hbox{\Large \boldmath$)$} \,\;
  H_{\rpp}^{hv} \6^{\prime\prime} H_{\rpp}^{ev} \,\d_{j_{\rpp}^e}
  \quad . \quad
\eea % A10
Obviously, under way from \eq{A9} to \eq{A10}, the translation operator 
has changed its place. This was justified by partial integration under 
$\int^\prime\,$. The right end of \eq{A10} that follows the round bracket
 can be dealt with as $\lk \6^{\prime\prime} + j_{\rpp} \rki^{eh} \,\d_{j_{\rpp}^e}\,$.

We now encounter a surprizing problem in forming the limit $\e \to 0$
(it is the \glqq harmless\grqq\ $\e\,$). In \eq{A10} the second term 
in the round bracket consists of three factors (counting
$\ov{\cl G }^{au}_{r \rp}$ as two, see \eq{2.2}, and including
only $\ov G_{r \rp}$ next). The product of
the first two reads
\be{A11}
  \cl P \,\gll\, {\e^2/\pi \0 
   \lk (z^{\prime\prime} -z^\prime)\,(\ov z^{\prime\prime} - \ov z^\prime)
    + \e^2 \rki^2 } \; \cdot\, {(\ov z^{\prime\prime} -\ov z^\prime )/\pi \0
    (z -z^\prime )\,(\ov z^{\prime\prime} -\ov z^\prime ) + \e^2 } \;\; . \;\;
\ee % A11
If (A) the limit is first performed in the first fraction, giving 
$\d \(\vc r^{\prime\prime} -\vc r^\prime\,\)$, $\,\cl P = 0\,$ is obtained.
But if (B) the limit is first done in the second fraction one
ends up with $\cl P = \d \(\vc r^{\prime\prime} -\vc r^\prime\) \cdot
{1 \0 \pi (z-z^\prime) }\,$. Who wins \anke (A) or (B) or none of\,?
To resolve this puzzle we integrate $\cl P$ over all $\vc r^{\prime\prime}$
and ignore the less important dependences on $\vc r^{\prime\prime}$ outside $\cl
P\,$\,:
\be{A12} \cl Q \,\gll\, \int\! d^2 r^{\prime\prime} \; \cl P \,=\,
 \int\! d^2 r^{\prime\prime} \; {\e^2/\pi \0 
   \lk r^{\prime\prime\;2} + \e^2 \rki^2 } \; 
   \cdot\, {\ov z^{\prime\prime}/\pi \0
    (z -z^\prime )\, \ov z^{\prime\prime} + \e^2 } \;\; . \;\;
\ee % A12
With $(z -z^\prime) = \o {\rm e}^{i \b}\,$ ($\o$ real)\,, $\ov z^{\prime\prime}
= \rho \,{\rm e}^{i (\ph - \b)}\,$ (\,hence 
$z^{\prime\prime} = \rho \,{\rm e}^{- i (\ph - \b)}\,$) and with 
the notation $\schl \e \gll \e /(\o \rho)\,$ the integral turns into
\be{A13}
 \!\!\!\cl Q = {1 \0 \pi^2 \o {\rm e}^{i \b}} \int_0^\infty\!\!\!d\rho\;
  {\rho \0 \lk \rho^2 +1\rki^2}\,\cl J \;\, {\rm with} \;\, \cl J 
   = \int_{-\pi}^\pi \!\!\! d\ph {1 \0 1 + \schl \e {\rm e}^{-i\ph}} = \ldots = 
   2\pi \theta\( \o \rho - \e \) \;\,
\ee % A13
and $\theta$ the step function. In the final result for $\cl Q\,$ the limit
$\e \to 0$ is unambiguous\,:
\be{A14}
 \cl Q \,=\, {1/\pi \0 z-z^\prime} \int_{\e/\o}^\infty\!d\rho \; {2\,\rho \0
 \lk \rho^2 +1\rki^2} \;\; \longrightarrow \;\;  {1 \0 \pi \,(z-z^\prime)} 
 \quad . \quad
\ee % A14
Version (B) wins. We conclude that $\cl P$ can be replaced by
${1 \0 \pi \,(z-z^\prime)} \, \d\(\vc r^{\prime\prime} -\vc r^\prime\)\,$.

The Delta function is now common to both terms in the round bracket of
\eq{A10} and helps performing $\int^{\prime\prime}\,$. The intermediate
result for $\cl T_1$ so far reached is
\bea{A15}
 \cl T_1 &=& {i\0 N}\!\int\!\!\int^\prime  
   {e_{r \rp} \0 \pi (\vc r -\vc r^\prime)^2}\; f^{cuh}\, \cdot \nonu \\ 
  & & \(  \, e_{r \rp}  \rule[-.2mm]{0pt}{5mm}
  H_r^{ab} H_{z^\prime \ov z}^{cb} H_{z {\ov z}^\prime}^{ad} H_{\rp}^{ud}
  -H_r^{ab} H_{z^\prime \ov z}^{cb} H_r^{ad} H_{z^\prime \ov z}^{ud}
  \;\)  \lk \6^{\prime} + j_{\rp} \rki^{eh} \d_{j_{\rp}^e} \quad . \quad
\eea % A15
Note that the second term in the round bracket reduces to $\d^{cu}\,$.
So it vanishes in contact with $f^{cuh}\,$. By partial integration
the square bracket can be placed as $\lk -\6^{\prime} + j_{\rp} \rki^{eh}$ 
at the beginning of the integrand. Then of course $-\6^\prime$ does
no more act on $\d_{j_{\rp}^e}\,$. Finally we interchange primed
and unprimed variables. Remembering 
$\cl T_1 = \!\int\! \o_r^{\heartsuit a}\,\d_{j_r^a}\,$ the object
$\o_r^{\heartsuit a}$ can now be read off to be \eq{2.8}
in the main text.

\rule[.5mm]{3cm}{.9mm}\, \eq{2.9} : \\[1mm]
To learn about $\cl T_2 = \int \o^{\diamondsuit a} \d_{j_r^a}\,$
the functional derivative $\ov{p}_{\rp}^u$ in \eq{2.7} has to be
placed in front of $H_{\rpp}^{ev}\,$. Using \eq{2.3} and \eq{A5}  
we have
\bea{A16}
 \ov{p}_{\rp}^u H_{\rpp}^{ev}\, &=&
  -\ov \6'\,( M_{\rp}^\dagger )^{ud} \d_{A^{d *}(\vcsm r')} H_{\rpp}^{ev}\,
   \nonu \\
   &=& -\ov \6'\,( M_{\rp}^\dagger )^{ud}
    {\ov G}_{\rpp \rp} (M_{\rp}^\dagger)^{gd}\,f^{egh}\, H_{\rpp}^{hv} 
   = - \d (\vc r^{\prime\prime} - \vc r^\prime) \,f^{euh}\, H_{\rpp}^{hv} 
   \quad \quad
\eea % A16
  and consequently
\bea{A17}
   \cl T_2 \glo{7} -{\pi\0 N}\int\!\int^\prime\!\int^{\prime\prime} H_r^{ab}
   \; \ov{\cl G }^{au}_{r \rp} {\cl G }^{bv}_{r \rpp} \; 
    i \6^{\prime\prime} \d (\vc r^{\prime\prime} - \vc r^\prime) 
    \,f^{euh}\, H_{\rpp}^{hv} \d_{j_{\rpp}^e} \nonu \\
   \gluo{5}{6}  
    { i \pi\0 N}\int\!\int^\prime\!  H_r^{ab}
   \; \ov{\cl G }^{au}_{r \rp} \( \6^\prime {\cl G }^{bv}_{r \rp} \)  
     \,f^{euh}\, H_{\rp}^{hv} \d_{j_{\rp}^e} 
    \;\, \hbox{\small or by just changing notation} \nonu \\[-4mm]
     & & \hspace*{3cm}
      \vc r \to \vc r^{\prime\prime} \,,\, \vc r^\prime \to \vc r \,,\,
      v \to c \,,\, h \to v \,,\, a \to d \,,\, e \to a \, : \nonu \\[-4mm]
  \glu{9} \int \lk {i\0 N} f^{auv} (-\pi ) \!\int^{\prime\prime}\!
      \ov{\cl G }^{du}_{\rpp r} H_{\rpp}^{db} \( \6 {\cl G }^{bc}_{\rpp r}\)  
      \, (H_r^{-1})^{cv} \rk \d_{j_r^a} \quad . \quad 
\eea % A17
The square bracket is $\o_r^{\diamondsuit a}\,$ indeed, see \eq{2.9}.

\rule[.5mm]{3cm}{.9mm}\, \eq{2.10} : \\[1mm]
According to the text below \eq{2.7} for the third part 
$\cl T_3 = \!\int\!\!\!\int^{\prime}\!\O_{r r'}^{a b}
\,\d_{j_r^a}\d_{j_{\rp}^b}$ we have to place $\ov{p}_{\rp}^u$ in front
of $\d_{j_{\rpp}^e} \psi\!\lk j\rk$ ($\glr \chi\!\lk j \rk$ for brevity).
\be{A18}
 \ov{p}_{\rp}^u \chi = -\ov \6'\, ( M_{\rp}^\dagger )^{ud} 
    \d_{A^{d *}(\vcsm r')} \chi = 
     -\ov \6'\, ( M_{\rp}^\dagger )^{ud} \int^{\prime\prime} 
     \( \d_{A^{d *}(\vcsm r')} j_{\rpp}^c \)
     \d_{ j_{\rpp}^c} \chi  
\ee % A18 
The round bracket is evaluated via $j_{\rpp}^c = 2 \Sp \( T^c
(\6^{\prime\prime} H_\rpp ) H_{\rpp}^{-1}\)$ and the right half of
\eq{A6}\,. Again abbreviating $M_{\rp}^\dagger T^d M_{\rp}^{\dagger -1}$
by $D$ it follows that 
\bea{A19}
 \d_{A^{d *}(\vcsm r')}\,j_{\rpp}^c \glo{6}
    -i ( \6^{\prime\prime} \ov G_{\rpp \rp}) 2 \Sp(T^c D)
    -i \ov G_{\rpp \rp} 2 \Sp( \lk T^c , D\rk j_\rpp ) \qquad \nonu \\
  \gluo{6}{6} 
    -i (\6^{\prime\prime} \ov G_{\rpp \rp}) (M_{\rp}^\dagger)^{cd}
    -i \ov G_{\rpp \rp} 2 \Sp( \lk T^c , D\rk T^p )\, 2 \Sp (T^p j_\rpp )
     \qquad \nonu \\
  \glu{6} 
     -i (M_{\rp}^\dagger)^{cd} \6^{\prime\prime} \ov G_{\rpp \rp}
     + (M_{\rp}^\dagger)^{qd} \,\ov G_{\rpp \rp} f^{cqp} j_{\rpp}^p
    \quad . \quad 
\eea % A19
Using \eq{A19} in \eq{A18} we obtain
\bea{A20}  
  \ov{p}_{\rp}^u \chi &=&  \int^{\prime\prime} 
  \( - i \d^{uc} \6^{\prime\prime} \d (\vc r^{\prime\prime} - \vc r^\prime )
     + \d (\vc r^{\prime\prime} - \vc r^\prime ) f^{cup} j_{\rpp}^p \)
     \d_{ j_{\rpp}^c} \chi \qquad \nonu \\[1mm]
    &=& 
   i \( \,\6^\prime - j_{\rp}\, \)^{uc} \,\d_{ j_{\rp}^c} \chi
   \quad \quad
\eea % A20
to be inserted in \eq{2.7}\,:
\bea{A21}
  \cl T_3 \glo{7}{\pi\0 N}\int\!\int^\prime\!\int^{\prime\prime} H_r^{ab}
  \; \ov{\cl G }^{au}_{r \rp} {\cl G }^{bv}_{r \rpp} \; 
    i \6^{\prime\prime} H_{\rpp}^{ev}
   i \( \,\6^\prime - j_{\rp} \)^{uc} 
      \,\d_{ j_{\rp}^c} \d_{j_{\rpp}^e} \;\; , \quad \nonu \\[-4mm]
   & & \hspace*{2cm} \hbox{\small partial integrations and
          $\vc r , \vc r^\prime , \vc r^{\prime\prime} \,\to\,
          \vc r^{\prime\prime} , \vc r , \vc r^\prime\;$:} \nonu \\[-3mm]
   \glu{9} 
   \int\!\!\int^\prime {1\0 N} \( \,\6 - j_r \)^{cu}   
   (-\pi)\!\int^{\prime\prime}
   \!\ov{\cl G }^{au}_{\rpp r} H_{\rpp}^{ab} 
     \( \6^{\prime} {\cl G }^{bv}_{\rpp \rp}\)
      (H_{\rp}^{-1})^{ve} \,\d_{ j_r^c} \d_{j_{\rp}^e} \;\; . \quad
\eea % A21
Now \eq{2.10} can be read off from \eq{A21}. End of 
\cite{kkn}--recapitulations.

\let\dq=\thq \renewcommand{\theequation}{B.\dq}
                           \setcounter{equation}{0}

\parag{Appendix B : \,\hbox{\small Evaluation of the
        integral {\boldmath$I_{p,q}(\ep,\s)\,$}, \eq{3.15}}}

At first we drop the double primes in \eq{3.15} and
set $z = r e^{-i\ph}$, $\ov z = r e^{i\ph}\,$: 
\be{B1}
  I_{p,q}(\ep,\s) = {1\0 \pi} \!\int_0^\infty \! dr \; r\, e^{-r^2/\ep}\,
  r^{p+q-1} \!\int\!\! d\ph \,{ (e^{i\ph})^{q-p-1} \0
  r e^{-i\ph} + \s } \quad . \quad
\ee % B1
Next we write $\s = |\s| e^{-i\chi}$ and
\be{B2} 
  I_{p,q}(\ep,\s) = {1\0 \pi} \!\int_0^\infty \!\!\! dr \; r\, e^{-r^2/\ep}\,
  r^{p+q-1} \s^{p-q} E_{q-p} \qquad \hbox{with} \qquad 
\ee % B2
\be{B3}  
    E_n \,\gll \int\!\! d\ph \,{ |\s|^n (e^{-i\chi})^n (e^{i\ph})^{n-1}
    \0 r e^{-i\ph} + |\s| e^{-i\chi} } \;=\; \!\int\!\! d\ph \,
    { ( |\s| e^{i\ph} )^n \0 r + |\s| e^{i\ph} } \quad , \quad
\ee % B3
where the last step rests on the shift $\ph \to \ph + \chi\,$.
Writing the numerator as \\
$( |\s| e^{i\ph} )^{n-1} (|\s|e^{i\ph} + r -r)$
one easily derives the recurrence relation
\be{B4}
   E_n = \d_{n,1}\, 2\pi - r E_{n-1} \quad , \quad 
   E_0 = \int\! d\ph {1 \0 r + |\s|e^{i\ph} } = {2\pi \0 r} 
    \theta(r-|\s|) \quad . \quad
\ee % B4
$\theta$ is the step function, cf.\,\eq{A13}\,. The solution to the problem 
\eq{B4} is 
\be{B5}
   E_n = \lower .8mm\hbox{\LARGE \{ }\!\!\matrix{
     \ \,(-r)^{n-1} \,2\pi \,\theta(|\s|-r)\ \ \hbox{for}\ \ n \ge 1 \cr
       - (-r)^{n-1} \,2\pi \,\theta(r-|\s|)\ \ \hbox{for}\ \ n \le 0 \cr }   
       \quad . \quad  
\ee % B5
With \eq{B5} at $n=q-p$, substituting $t= r^2$ and using 
  $|\s|^2=\s \ov \s$ from
\eq{3.13}\,, \eq{B2} turns into
\bea{B6}
   I_{q\le p} = (-\s)^{p-q} \int_{\s \ov\s}^\infty \! dt \; t^{q-1} e^{-t/\ep}
   \quad \hbox{and} \quad
   I_{q>p} = -(-\s)^{p-q} \int_0^{\s \ov \s} \! dt \; t^{q-1} e^{-t/\ep}
   \quad . \quad
\eea
Through $\int_0^{\s \ov \s} = \int_0^\infty - 
       \int_{\s \ov \s}^\infty$ one arrives at
\eq{3.17} in the main text. 

\let\dq=\thq \renewcommand{\theequation}{C.\dq}
                           \setcounter{equation}{0}

\parag{Appendix C : \,\hbox{\small Example of a MAPLE--file}}

In evaluating $\cl T_3\, Q_1 $ the five steps described
below \eq{4.14} lead to 

  \hspace*{1mm}
   aa9 := \\ 
 +2$*$ab$*$L23 \\
 +4$*$abj$*$L13 \\
 +6$*$abtj$*$L12 \\
  \hspace*{.5mm}
 -6$*$abttj$*$L11 \\
  \hspace*{.5mm}
 -4$*$abtttj$*$L10 \\[-6mm]

 +4$*$abdj$*$L03 \\
 +6$*$abdtj$*$L02 \\
  \hspace*{.5mm}
 -6$*$abdttj$*$L01 \\
  \hspace*{.5mm}
 -4$*$abdtttj$*$L00 \\[-6mm]

 +2$*$ab$*$jL13  \\
 +4$*$abj$*$jL03  \\
 +6$*$abtj$*$jL02  \\
  \hspace*{.5mm}
 -6$*$abttj$*$jL01 \\
  \hspace*{.5mm}
 -4$*$abtttj$*$jL00 :

  \vskip -8.2cm \hspace*{5cm} \parbox[t]{8cm}{
  Of course $a$ stands for $f^a$ and $b$ for $f^b\,$. Read \\
  the factor containing them as an adjoint trace, hence especially 
  ab$ = (ab)^{cc} = f^{cad}f^{dbc}$ \\
  $ = -N \d^{ab}$. But the factors containing L carry 
  indices$^{ab}$. Hence e.g. jL02 = $(j * L_{02} )^{ab}\,$.
  \\[3mm]
  dttj is $\6 \,\ov\6^2 j\,$, hence \,abdttj$ = f^{cad}f^{dbe}
  \,\6 \,\ov\6^2 j^{ec}$. \\
  Using $f^{cad}f^{dbe}f^{euc} = -(N/2)\,f^{abu}$ this turns 
  into $(N/2)\,\6\, \ov{\ov \jmath}{}^{\,ab}$. 
  \\[3mm]
  By the cyclic invariance of trace a can always be moved to the
  left end inside $(\;\;\;)^{cc}$. By inversion of the content of 
  $(\;\;\;)^{cc}$ (the factors are anti--symmetric) also $b$
  can be placed near to a. But to have always ab at the left
  end is a speciality of the example $\cl T Q_1\,$ choosen. } \\

\let\dq=\thq \renewcommand{\theequation}{D.\dq}
                           \setcounter{equation}{0}

\parag{Appendix D : \,\hbox{\small List of the \boldmath$L$'s which
                      occur in Appendix C}}

How $L$--coefficients look like. Here they include only the first
power of $\ep\,$ (to limit the length of this paper). Each term
carries indices$^{ab}$. $\,...j\;\!${something} means $\,...j*$something. 
\bea{D1} 
L00 &=& +j \,\; +\ep*\Big( +3/8*dtj +3/8*tjj -3/8*jtj 
     \;\Big)   
     \\[3mm]
L01 &=& +1/2*{1\0\ep} +1/4*tj \nonu \\
    & & +\ep*\Big(\; +7/16*dttj +7/16*ttjj -7/16*jttj +1/8*tjtj 
     \;\Big)
     \\[3mm]
L02 &=&  +\ep*\Big(\; +15/32*dtttj +15/32*tttjj -15/32*jtttj \nonu \\
  & & \hspace*{14mm} +17/32*ttjtj -7/32*tjttj
    \;\Big)
    \\[3mm]
L03 &=& -11/48*tttj \,\; +\ep*\Big(\; +31/64*dttttj +31/64*ttttjj \nonu \\
    & &  +1*tttjtj +3/16*ttjttj  -5/8*tjtttj -31/64*jttttj 
     \;\Big)
     \\[3mm]
L10 &=& +1/2*dj -1/2*jj \,\; +\ep*\Big(\; +7/24*ddtj +5/24*dtjj -7/24*djtj 
          \nonu \\
    & & +7/24*tjdj -7/12*jdtj -1/12*tjjj -5/24*jtjj +7/24*jjtj 
     \;\Big)
     \\[3mm]
L11 &=& 
   -1/2*{j\0 \ep} -1/4*tjj \,\; +\ep*\Big(\;
 +31/96*ddttj
  +5/24*dttjj \nonu \\ & &
  -1/12*dtjtj
 -31/96*djttj
 +31/96*ttjdj
  -5/48*tjdtj \nonu \\ & &
 -31/48*jdttj
 -11/96*ttjjj
  -5/24*jttjj 
 +31/96*jjttj \nonu \\ & &
 -11/48*tjtjj
  +1/48*tjjtj
  +1/12*jtjtj
   \;\Big) 
   \\[3mm]
L12 &=& -1/4*{1\0\ep^2} -1/4*{tj\0\ep} -11/32*dttj \nonu \\
    & & -11/32*ttjj +11/32*jttj -3/16*tjtj \nonu \\
 +\ep*\Big( \hspace*{-8mm} 
     & &
   +1/3*ddtttj
 +19/96*dtttjj
  +1/16*dttjtj
  -7/16*dtjttj
   -1/3*djtttj \nonu \\ & &
   +1/3*tttjdj
   +1/8*ttjdtj
   -5/8*tjdttj
   -2/3*jdtttj
 -13/96*tttjjj \nonu \\ & &
 -19/96*jtttjj
   +1/3*jjtttj
 -13/32*ttjtjj
  -1/16*ttjjtj
 -13/32*tjttjj \hspace*{-12mm} \nonu \\ & &
  -1/16*jttjtj
  +3/16*tjjttj
  +7/16*jtjttj
   -1/8*tjtjtj
   \;\Big) 
   \\[3mm]
 L13 &=& -1/16*{ttj\0\ep} -61/96*dtttj -13/32*tttjj +61/96*jtttj
         -17/32*ttjtj \nonu \\ 
     & & +7/32*tjttj \nonu \\
 +\ep*\Big( \hspace*{-8mm} 
     & & 
 +43/128*ddttttj
   +3/16*dttttjj
   +7/32*dtttjtj
  -15/32*dttjttj \nonu \\ & &
  -19/16*dtjtttj
  -43/64*djttttj
  -43/64*jdttttj
  -39/32*tjdtttj \nonu \\ & &
  -45/64*ttjdttj
  +13/32*tttjdtj
 +37/128*ttttjdj
 -19/128*ttttjjj \nonu \\ & &
   -3/16*jttttjj
 +43/128*jjttttj
  -19/32*tttjtjj
   -3/16*tttjjtj \nonu \\ & &
  -19/32*tjtttjj
   -7/32*jtttjtj
  +13/32*tjjtttj
  +13/16*jtjtttj \nonu \\ & &
  -57/64*ttjttjj
  +15/64*ttjjttj
  +15/32*jttjttj
   -9/16*ttjtjtj \nonu \\ & &
   -9/16*tjttjtj
  +15/32*tjtjttj
   \;\Big) 
   \\[3mm]
 L23 &=& 
 +1/4*{1\0 \ep^3}
 +3/8*{tj\0 \ep^2}
 +3/8*{dttj\0 \ep}
 +1/2*{ttjj\0 \ep}
 -3/8*{jttj\0 \ep}
 +3/8*{tjtj\0 \ep} \nonu \\ & &
 -79/96*ddtttj
   -3/8*dtttjj
   -1/4*dttjtj
 +25/32*dtjttj
 +79/96*djtttj  \nonu \\ & &
 +79/48*jdtttj
 +35/32*tjdttj
 -13/32*ttjdtj
 -19/32*tttjdj
  +7/32*tttjjj \hspace*{-12mm}  \nonu \\ & &
   +3/8*jtttjj
 -79/96*jjtttj
 +21/32*ttjtjj
  +5/32*ttjjtj
   +1/4*jttjtj \nonu \\ & &
 +21/32*tjttjj
  -5/16*tjjttj
  25/32*jtjttj
  +5/16*tjtjtj \nonu \\
  +\ep*\Big(\; \hspace{-8mm} & &
 +129/512*dddttttj
  +43/512*ddttttjj
   -1/128*ddtttjtj
 -139/256*ddttjttj \hspace*{-8mm} \nonu \\ & &
 -275/384*ddtjtttj
 -129/512*ddjttttj
 -387/512*jddttttj
 -199/128*tjddtttj \hspace*{-8mm} \nonu \\ & &
 -309/256*ttjddttj
   +9/128*tttjddtj
 +129/512*ttttjddj
 +139/512*dttttjdj  \hspace*{-8mm} \nonu \\ & &
   +3/128*dtttjdtj
 -399/256*dttjdttj
 -261/128*dtjdtttj
 -387/512*djdttttj \hspace*{-8mm} \nonu \\ & &
  -53/512*dttttjjj
  -53/128*dtttjtjj
   -5/128*dtttjjtj
 +121/256*dttjjttj \hspace*{-8mm} \nonu \\ & &
 -159/256*dttjttjj
  -15/128*dttjtjtj
  -53/128*dtjtttjj
 +233/384*dtjjtttj \hspace*{-8mm} \nonu \\ & &
  -15/128*dtjttjtj
 +121/128*dtjtjttj
  -43/512*djttttjj
   +1/128*djtttjtj \hspace*{-8mm} \nonu \\ & &
 +139/256*djttjttj
 +275/384*djtjtttj
 +129/512*djjttttj
  -43/256*jdttttjj \hspace*{-8mm} \nonu \\ & &
    +1/64*jdtttjtj
 +139/128*jdttjttj
 +275/192*jdtjtttj
 +129/256*jdjttttj \hspace*{-8mm} \nonu \\ & &
   -43/64*tjdtttjj
   -15/64*tjdttjtj
   +85/64*tjdtjttj
 +161/192*tjdjtttj \hspace*{-8mm} \nonu \\ & &
 -129/128*ttjdttjj
   -15/64*ttjdtjtj
  +85/128*ttjdjttj
   -43/64*tttjdtjj \hspace*{-8mm} \nonu \\ & &
    -5/64*tttjdjtj
  -43/256*ttttjdjj
 +387/512*jjdttttj
 +137/128*tjjdtttj \hspace*{-8mm} \nonu \\ & &
 +261/128*jtjdtttj
 +219/256*ttjjdttj
 +219/128*tjtjdttj
 +399/256*jttjdttj \hspace*{-8mm} \nonu \\ & &
  -15/128*tttjjdtj
  -45/128*ttjtjdtj
  -45/128*tjttjdtj
   -3/128*jtttjdtj \hspace*{-8mm} \nonu \\ & &
 -119/512*ttttjjdj
 -119/128*tttjtjdj
 -357/256*ttjttjdj 
 -119/128*tjtttjdj \hspace*{-8mm} \nonu \\ & &
 -139/512*jttttjdj
  +33/512*ttttjjjj
  +53/512*jttttjjj
  +43/512*jjttttjj \hspace*{-8mm} \nonu \\ & &
 -129/512*jjjttttj
  +33/128*tttjtjjj
  +33/128*tttjjtjj
   +5/128*tttjjjtj \hspace*{-8mm} \nonu \\ & &
   +5/128*jtttjjtj
  +53/128*jtttjtjj
  +33/128*tjtttjjj
  +33/128*tjjtttjj \hspace*{-8mm} \nonu \\ & &
  +53/128*jtjtttjj
   -1/128*jjtttjtj
  -89/384*tjjjtttj
 -233/384*jtjjtttj \hspace*{-8mm}  \nonu \\ & &
 -275/384*jjtjtttj
  +99/256*ttjttjjj
  +99/256*ttjjttjj
  -49/256*ttjjjttj \hspace*{-8mm} \nonu \\ & &
 +159/256*jttjttjj
 -121/256*jttjjttj
 -139/256*jjttjttj
  +99/128*ttjtjtjj \hspace*{-8mm} \nonu \\ & &
  +15/128*ttjtjjtj
  +15/128*ttjjtjtj
  +99/128*tjttjtjj
  +15/128*tjttjjtj \hspace*{-8mm} \nonu \\ & &
  +15/128*jttjtjtj
  +99/128*tjtjttjj
  +15/128*tjjttjtj
  +15/128*jtjttjtj \hspace*{-8mm} \nonu \\ & &
  -49/128*tjtjjttj
  -49/128*tjjtjttj \nonu \\ & &
 -121/128*jtjtjttj
   +15/64*tjtjtjtj
      \;\Big)
\eea

\end{document}